\documentclass[canadian,english,nofootinbib,notitlepage,aip,jcp,reprint,citeautoscript]{revtex4-1}
\usepackage[T1]{fontenc}
\usepackage[latin9]{inputenc}
\usepackage{babel}
\usepackage{amsmath}
\usepackage{amssymb}
\usepackage{graphicx}
\usepackage[unicode=true,pdfusetitle,
 bookmarks=true,bookmarksnumbered=false,bookmarksopen=false,
 breaklinks=false,pdfborder={0 0 1},backref=false,colorlinks=false]
 {hyperref}

\makeatletter

\providecommand{\tabularnewline}{\\}

\usepackage{braket,listings,xcolor}

\makeatother

\begin{document}
\selectlanguage{canadian}%
\global\long\def\sbn#1#2{\{#1\:|\:#2\}}%

\global\long\def\norm#1{\Vert#1\Vert}%

\global\long\def\abs#1{|#1|}%

\global\long\def\d{\mathrm{d}}%

\global\long\def\f#1#2#3{#1:\ #2\rightarrow#3}%

\global\long\def\k#1{\mathop{\mathrm{#1}}\nolimits}%

\global\long\def\braket#1#2{\Braket{#1|#2}}%

\global\long\def\bra#1{\Bra{#1}}%

\global\long\def\ket#1{\Ket{#1}}%

\global\long\def\C#1{\cancel{#1}}%

\global\long\def\BC#1{\bcancel{#1}}%

\global\long\def\u#1{\mathrm{\:#1}}%

\global\long\def\vac{\ket{\underbar{0}}}%

\global\long\def\vacb{\bra{\underbar{0}}}%

\global\long\def\dag#1{#1^{\dagger}}%

\global\long\def\expval#1{\langle#1\rangle}%

\global\long\def\cvec#1{\boldsymbol{#1}}%

\renewcommand{\Im}{\text{Im}}

\providecommand{\Tr}{\mathrm{Tr}}

\selectlanguage{english}%
\providecommand{\diffbegin}{\color{black}}

\providecommand{\diffend}{\color{black}}

\selectlanguage{canadian}%
\renewcommand*{\thefootnote}{\roman{footnote}}

\global\long\def\alg{\text{UF}^{2}}%

\selectlanguage{english}%
 
\title{Numerical Method for Nonlinear Optical Spectroscopies: \\Ultrafast
Ultrafast Spectroscopy}
\author{Peter A.~Rose}
\affiliation{Department of Physics, University of Ottawa, Ottawa, ON, Canada}

\author{Jacob J.~Krich}
\affiliation{Department of Physics, University of Ottawa, Ottawa, ON, Canada}
\affiliation{School of Electrical Engineering and Computer Science, University
of Ottawa, Ottawa, ON, Canada}
\begin{abstract}
We outline a novel numerical method, called Ultrafast Ultrafast ($\alg$)
spectroscopy, for calculating the $n^{\text{th}}$-order wavepackets
required for calculating $n$-wave mixing signals. The method is simple
to implement, and we demonstrate that it is computationally more efficient
than other methods in a wide range of use cases. Resulting spectra
are identical to those calculated using the standard response function
formalism but with increased efficiency. The computational speed-ups
of $\alg$ come from (a) non-perturbative and costless propagation
of the system time-evolution (b) numerical propagation only at times
when perturbative optical pulses are non-zero and (c) use of the fast
Fourier transform convolution algorithm for efficient numerical propagation.
The simplicity of this formalism allows us to write a simple software
package that is as easy to use and understand as the Feynman diagrams
that \diffbegin organize the understanding of $n$-wave mixing processes. \diffend
\end{abstract}
\maketitle

\section{Introduction}

Ultrafast nonlinear optical spectroscopies, in the perturbative light-matter
limit, are powerful tools for elucidating details about the electronic
structure and ultrafast dynamics of optically active systems. Interpreting
such spectra often requires understanding what signals would be produced
by a range of parametrized system Hamiltonians; one then seeks the
best agreement with experimental results by varying system parameters
such as energy levels, couplings, dephasing rates, and more. This
kind of fitting, as in Ref.~\onlinecite{Perdomo-Ortiz2012}, involves
repeatedly rederiving the spectroscopic signals as system parameters
change, which can be computationally expensive.

Simulations of nonlinear spectroscopies can be particularly challenging
when the optical pulse durations are similar to relevant timescales
in the system dynamics, especially when one must consider the pulses
overlapping in time. \diffbegin The response-function formalism, described further
below, provides a powerful way to understand nonlinear spectra and
can be computationally efficient when considering the limit of impulsive
optical pulses, those with durations shorter than any relevant system
dynamics \citep{Mukamel1999}. \diffend For spectroscopies that rely on varying
pulse durations, simulating those variations can be computationally
expensive, limiting the range of systems and pulses one can study
\citep{Johnson2014}.

\diffbegin The effects of finite pulse durations can be studied using perturbative
\citep{Engel1991,Meyer1999,Kato2001,Tsivlin2006,Cheng2007,Renziehausen2009,Tanimura2012,Yuen-Zhou2014,Johnson2014,Bell2015,Cina2016,Perlik2017,Smallwood2017,Do2017,Fetherolf2017}
and nonperturbative \citep{Seidner1995,Mancal2006,Seibt2009} methods
\citep{Domcke2007}. Except in special cases, where the dynamics can
be solved analytically \citep{Perlik2017,Smallwood2017,Johnson2014,Cina2016},
generic methods numerically solve the time-dependent Schrodinger equation
to determine the system response and spectroscopic observables. In
the frequently studied cases of electronic excitations coupled to
vibrations or optical cavities, without breaking of chemical bonds,
this integration is performed efficiently in a basis of electronic
and vibrational/optical excitations, with respect to which the system
Hamiltonian is generally highly sparse. The numerical integration
is frequently performed using Runge-Kutta (RK) methods \citep{Gelin2005,Gelin2005a,Gelin2009,Gelin2009a,Tanimura1997,Yan2017,Fetherolf2017},
with other frequently used quantum dynamics packages using Adams-Bashforth
\citep{Johansson2012} or Bulirsch-Stoer \citep{Beck2000,Tsivlin2006}
methods. An alternative approach, especially valuable in cases with
bond breaking, where a continuum of vibrational states become relevant,
uses a real- and Fourier-space pseudospectral representation of the
wavefunction, frequently using the split-operator method to propagate
the system dynamics \citep{Feit1982,Kosloff1983,Seibt2009,Renziehausen2009}. \diffend

There has recently been an increased interest in understanding the
role of finite-pulse-duration effects in nonlinear spectroscopies,
and there are several \diffbegin analytic methods for the propagation of the
states required for calculating 4-wave mixing signals using particular
shapes of finite pulses \citep{Perlik2017,Smallwood2017,Johnson2014}.
Each of these methods assumes a particular, idealized pulse shape,
such as a Gaussian \citep{Smallwood2017,Johnson2014} or Lorentzian
\citep{Perlik2017} profile, and some include pulse overlap effects.
These methods require that the eigenvalues and eigenvectors of the
unperturbed system are known, allowing use of analytic results for
the time-dependent perturbation theory (TDPT), which can be summed
over all required states. \diffend

We present a complementary numerical technique we call Ultrafast Ultrafast
($\alg$) spectroscopy, \diffbegin which combines features of the analytic and
full numerical integration approaches. $\alg$ assumes that the eigenvalues
and eigenvectors of the unperturbed system are known but solves the
integrals arising from TDPT numerically and can therefore treat arbitrary
pulse shapes. \diffend $\alg$ works for arbitrary perturbative order $n$,
including pulse-overlap effects. It is a fast implementation of the
standard results in perturbative nonlinear spectroscopy in the electric
dipole approximation\citep{Mukamel1999}. The version we present here
is for closed systems.\footnote{$\alg$ can be extended to work with open systems, but we do not do
so here.} It assumes that the system Hamiltonian is time-independent and has
a finite relevant eigenbasis. When the pulses are finite, we perform
the time propagation using the convolution theorem and the fast Fourier
transform (FFT) and therefore benefit from the speed of the FFT. Working
in the energy eigenbasis allows non-perturbative and costless time
evolution of the system at times when the optical pulses are negligible,
and $\alg$ also solves the TDPT integrals during the pulses with
such a small cost that they may not be computationally limiting
even for reasonably large system Hamiltonians. These speed-ups give
a dramatic improvement over algorithms that involve numerical integration
of the wavefunctions or density matrices for each time-step.

\diffbegin In many cases, the limiting step of $\alg$ is evaluating the expectation value of
the dipole operator after the pulses have passed. The computational
cost of $\alg$ is dominated by the cost of matrix-vector multiplication,
and so benefits from all the speed of linear algebra optimization
on modern computers. As it relies on diagonalizing the Hamiltonian,
it is clear that at some system size $\alg$ will cease to be a competitive
technique.

Conventional wisdom holds that diagonalization of the system Hamiltonian
is too expensive to be worthwhile for prediction of nonlinear spectroscopies
\citep{Domcke2007}. Especially for sparse matrices, as occur in e.g.,
vibronic systems, full or partial diagonalizations to attain the relevant
eigenvalues and eigenvectors can be relatively efficient \citep{Calvetti1994};
more importantly, if one desires to study 2D spectra with a large
number of different pulse delays, the diagonalization cost must be
paid only once. Further, one can consider the effects of different
pulse shapes, durations, and polarizations; various dipole coupling
matrices; and thermal or rotational averaging, all without needing
to rediagonalize. In many such cases, the diagonalization cost can
be insignificant compared to the cost of generating the spectra, even
for relatively large systems. \diffend

In Section~\ref{sec:The-Algorithm} we derive the mathematical framework
of $\alg$, show how it is used, and describe some of the techniques
that increase its efficiency. As with any perturbative technique,
it is most efficient where the rotating wave approximation and phase
matching can be assumed (see Section~\ref{subsec:RWA-PM}), but it
does not require these assumptions. \diffbegin In Section~\ref{sec:Computational-Advantage}
we compare the computational cost of $\alg$ to two alternatives:
a Runge-Kutta-based direct propagation method, and a split-operator
pseudospectral method, where we focus on example vibronic Hamiltonians.
For sufficiently large systems, the direct propagation methods become
more efficient than $\alg$, but we show that for Hamiltonians with
dimensions up to $O(10^{4})$, which includes a large number of widely
studied cases, $\alg$ outperforms the direct propagation methods.
In the frequently studied case of systems with dimension less than
100, $\alg$ can be two orders of magnitude faster than direct propagation
methods. We demonstrate how $\alg$ enables computationally efficient
studies of the effects of varying system parameters and optical pulse
shapes in Section~\ref{sec:Example}, where we show transient absorption
(TA) spectra for a system with a Hamiltonian of dimension 28 and demonstrate
how $\alg$ can be used to perform rapid parameter sweeps. \diffend

$\alg$ is not only efficient but also easy and intuitive to use.
The method is built around single-sided Feynman diagrams and their
double-sided counterparts, which describe the perturbative pathways
contributing to desired spectroscopic signals. Using the method is
a simple process of translating a desired Feynman diagram into a set
of iterative function calls. A python implementation of $\alg$ is
available for download at https://github.com/peterarose/ultrafastultrafast.
This code includes examples of how to translate Feynman diagrams into
the language of $\alg$ and an example implementation that calculates
TA spectra. \diffbegin It also contains a comparison Runge-Kutta integration
method, usable with the same convenient interface. \diffend

\section{Algorithm\label{sec:The-Algorithm}}

\subsection{Overview}

We begin with a Hamiltonian of the form
\begin{equation}
\hat{H}=\hat{H}_{0}+\hat{H}'(t),\label{eq:Htot}
\end{equation}
where the light-matter interaction in the electric-dipole approximation
is
\begin{equation}
\hat{H}'(t)=-\hat{\cvec{\mu}}\cdot\cvec E(t),\label{eq:Hprime}
\end{equation}
and is treated as a perturbation, where $\hat{\cvec{\mu}}$ and $\cvec E(t)$
are the dipole moment of the system and the external electric field,
respectively. Bold face symbols indicate Cartesian vectors. $\hat{H}_{0}$
describes the material system (e.g., a molecule or quantum dot) and
has eigenstates $\ket{\phi}$ and eigenvalues $\hbar\omega_{\phi}$,
which are assumed to be known (either analytically or numerically).
Therefore the unitary time-evolution operator $\hat{U}_{0}(t-t_{0})=e^{-i\hat{H}_{0}(t-t_{0})/\hbar}$
is known. We assume that the set of eigenstates $\left\{ \ket{\phi}\right\} $
is finite, and that all relevant wavefunctions in this problem can
be expressed using $N$ eigenfunctions as
\begin{equation}
\ket{\psi(t)}=\sum_{\phi=1}^{N}e^{-i\omega_{\phi}t}c_{\phi}(t)\ket{\phi}.\label{eq:psi}
\end{equation}
The electric dipole operator $\hat{\cvec{\mu}}$ must be known in
the eigenbasis of $\hat{H}_{0}$, where we define matrix elements
as
\[
\cvec{\mu}_{\phi\phi'}=\bra{\phi}\hat{\cvec{\mu}}\ket{\phi'}.
\]
Note that only $\omega_{\phi}$ and $\cvec{\mu}_{\phi\phi'}$ are
needed. The eigenfunctions themselves are unnecessary if $\cvec{\mu}_{\phi\phi'}$
can be calculated in some other way. 

We describe the electric field classically as a sum of pulses, where
each pulse is denoted by a lowercase letter starting from $a$. A
typical 4-wave mixing signal would be calculated by using up to 4
pulses. We write the electric field as a sum over $L$ pulses,
\[
\cvec E(t)=\sum_{i=a,b,\dots,L}\cvec e_{i}\varepsilon_{i}(t)+\cvec e_{i}^{*}\varepsilon_{i}^{*}(t),
\]
where $\boldsymbol{e}_{i}$ is the possibly complex polarization vector,
and the amplitude $\varepsilon_{i}$ of each pulse is defined with
envelope $A_{i}$, central frequency $\omega_{i}$, wavevector $\cvec k_{i}$,
and phase \diffbegin $\Theta_{i}$ as
\begin{equation}
\varepsilon_{i}(t)=A_{i}(t-t_{i})e^{-i\left[\omega_{i}(t-t_{i})-\cvec k_{i}\cdot\cvec r+\Theta_{i}\right]},\label{eq:pulse_shape}
\end{equation}
\diffend where $t_{i}$ is the arrival time of each pulse, and we define the
Fourier transform of the pulse as 
\[
\tilde{\varepsilon}_{i}(\omega)=\frac{1}{\sqrt{2\pi}}\int_{-\infty}^{\infty}\d t\varepsilon_{i}(t)e^{i\omega t}.
\]
Then the light-matter interaction is a sum over rotating ($\varepsilon_{i}$)
and counter-rotating ($\varepsilon_{i}^{*}$) terms. We express these
terms individually as 
\begin{align}
\hat{H}_{j^{(*)}}'(t) & =-\hat{\cvec{\mu}}\cdot\cvec e_{j}^{(*)}\varepsilon_{j}^{(*)}(t)\label{eq:Hj}
\end{align}
so that
\begin{equation}
\hat{H}'(t)=\sum_{i=a,b,...,L}\hat{H}'_{i}(t)+\hat{H}'_{i^{*}}(t).\label{eq:Hdecomposition}
\end{equation}

In general, 4-wave mixing signals are calculated from the third-order
perturbed density matrix \citep{Mukamel1999}. For closed systems,
as described here, we can decrease the computational complexity of
the problem by calculating the necessary perturbed wavefunctions up
to third-order. We expand the true time-dependent wavefunction using
the usual perturbative expansion 
\begin{equation}
\ket{\psi(t)}=\ket{\psi^{(0)}(t)}+\ket{\psi^{(1)}(t)}+\ket{\psi^{(2)}(t)}+...\label{eq:perturbative-expansion}
\end{equation}
where $\ket{\psi^{(0)}(t)}$ is the initial time-independent (ignoring
a trivial phase) state, and $\ket{\psi^{(n)}(t)}$ is proportional
to $\left(H'(t)\right)^{n}$. $\alg$ is a fast method for calculating
$\ket{\psi^{(n)}(t)}$. The total polarization field is
\begin{align*}
\cvec P(t) & =\bra{\psi(t)}\hat{\cvec{\mu}}\ket{\psi(t)}\\
 & =\cvec P^{(1)}(t)+\cvec P^{(2)}(t)+\cvec P^{(3)}(t)+...
\end{align*}
To calculate $\cvec P^{(n)}(t)$, we expand $\ket{\psi(t)}$ using
Eq.~\ref{eq:perturbative-expansion} and keep only terms proportional
to $\left(\hat{H}'(t)\right)^{n}$.\citep{PeterHamm2011} As an example,
the $3^{\text{rd}}$-order polarization field is 
\[
\cvec P^{(3)}(t)=\bra{\psi^{(3)}(t)}\hat{\cvec{\mu}}\ket{\psi^{(0)}}+\bra{\psi^{(2)}(t)}\hat{\cvec{\mu}}\ket{\psi^{(1)}}+c.c.
\]

From $\cvec P^{(3)}(t)$ any $3^{\text{rd}}$-order spectroscopic
signal can be determined. $\cvec P^{(3)}(t)$ is the material response
\diffbegin coupled to emitted radiation \diffend after three optical interactions. In
2D photon echo (2DPE) spectroscopy, the response is due to a single
interaction with each of 3 separate pulses, with arrival times $t_{a},t_{b},t_{c}$.
The delay times between pulses are $t_{b}-t_{a}=\tau$ and $t_{c}-t_{b}=T$.
The polarization field is therefore a function of $\tau$ and $T$:
$\cvec P^{(3)}(\tau,T;t)$. Typically the quantity of interest is
actually the Fourier transform partner
\[
\cvec P^{(3)}(\tau,T;\omega)=\frac{1}{\sqrt{2\pi}}\int_{-\infty}^{\infty}\d t\cvec P^{(3)}(\tau,T;t)e^{i\omega t}.
\]
The polarization field produces a \diffbegin radiating optical \diffend field, which can
be heterodyne detected using a fourth local oscillator (LO) pulse.
For the case of the 2DPE rephasing signal, the desired signal is calculated
as \citep{Mukamel1999}
\begin{equation}
S_{\text{2D}}^{(3)}(\tau,T,\omega)=\Im\left[\tilde{\varepsilon}_{LO}^{*}(\omega)\cvec e_{LO}^{*}\cdot\cvec P_{\text{2D}}^{(3)}(\tau,T;\omega)\right].\label{eq:3variableSignal}
\end{equation}

We derive $\alg$ for the classic example given in Eq.~\ref{eq:3variableSignal}.
In Section~\ref{sec:Example}, we show example TA spectra, which
arise due to the interaction of two pulses, a pump and a probe, separated
by the delay time $T$. In the TA case, the first two interactions
happen with the pump pulse, and therefore $\tau=0$. The third interaction
comes from the probe, which also acts as the local oscillator. TA
calculations are thus a function of only two variables:
\begin{equation}
S_{\text{TA}}^{(3)}(T,\omega)=\Im\left[\tilde{\varepsilon}_{\text{probe}}^{*}(\omega)\cvec e_{\text{probe}}^{*}\cdot\cvec P_{\text{TA}}^{(3)}(T,\omega)\right].\label{eq:TASignal}
\end{equation}

Returning to the more general case with $\tau$ varying, the canonical
formula for calculating the third-order time-dependent polarization
field is 
\begin{align}
P^{(3)}(\tau,T,t) & =\iiint dt_{3}dt_{2}dt_{1}E(t-t_{3})E(t-t_{3}-t_{2}+T)\nonumber \\
 & \times E(t-t_{3}-t_{2}-t_{1}+T+\tau)R(t_{1},t_{2},t_{3})\label{eq:Mukamelian}
\end{align}
where we follow Mukamel and suppress the polarization of the fields,
and $R(t_{1},t_{2},t_{3})$ is a material response function containing
all of the factors arising from $\hat{H}_{0}$ and $\hat{\cvec{\mu}}$
\citep{Mukamel1999}. This formulation separates the material properties
($\hat{H}_{0}$ and $\hat{\cvec{\mu}}$) from the shapes of the electric
fields. If the electric fields are impulsive (i.e., short compared
with any system timescale), then the convolutions in Eq.~\ref{eq:Mukamelian}
are trivial and spectral signals may be calculated directly from the
response function.

In this paper we focus on the case where the shape of the electric
field cannot be ignored, and therefore studying the response function
alone is insufficient to predict, interpret or understand spectroscopic
observables. The triple-nested convolutions of Eq.~\ref{eq:Mukamelian}
are so costly that they are rarely carried out. We present a fast
method for calculating the $n^{\text{th}}$-order wavefunctions, and
thence the desired polarization $\cvec P^{(n)}(t)$, with arbitrary
pulse shapes. The wavepackets can be studied in conjunction with the
signal, giving intuition for the underlying physics \citep{Heller1981}.

\subsection{Derivation\label{subsec:Derivation}}

We follow standard time-dependent perturbation theory by considering
that at time $t_{0}$ the system begins in a time-independent state
$\ket{\psi^{(0)}(t_{0})}$, i.e., an eigenstate of $\hat{H}_{0}$.
The perturbation $\hat{H}'(t)$ \diffbegin is zero before $t_{0}$ and \diffend produces
$\ket{\psi(t)}$ as in Eq.~\ref{eq:perturbative-expansion}, where
$\ket{\psi^{(n)}}$ has contributions proportional to $(\hat{H}')^{n}$.
Then the standard result has \citep{PeterHamm2011} 
\begin{align}
\ket{\psi^{(n)}(t)} & =-\frac{i}{\hbar}\int_{t_{0}}^{t}ds\hat{U}_{0}^{-1}(s-t)\hat{H}'(s)\ket{\psi^{(n-1)}(s)}.\label{eq:TDSE}
\end{align}
Since $\ket{\psi^{(0)}}$ is time-independent (up to a trivial phase),
we are free to send $t_{0}\rightarrow-\infty$. We also substitute
$t'=t-s$ to arrive at
\[
\ket{\psi^{(n)}(t)}=-\frac{i}{\hbar}\int_{0}^{\infty}dt'\hat{U}_{0}(t')\hat{H}'(t-t')\ket{\psi^{(n-1)}(t-t')}.
\]
Using the decomposition of $\hat{H}'(t)$ in Eq.~\ref{eq:Hdecomposition},
we define $\ket{\psi^{(n)}(t)}$ as a sum over $2L$ terms
\begin{align*}
\ket{\psi^{(n)}(t)} & =\sum_{j=a,b,\dots,L}\left(\hat{K}_{j}+\hat{K}_{j^{*}}\right)\ket{\psi^{(n-1)}(t)},
\end{align*}
where
\[
\hat{K}_{j^{(*)}}=-\frac{i}{\hbar}\int_{0}^{\infty}dt'\hat{U}_{0}(t')\hat{H}'_{j^{(*)}}(t-t').
\]

A small rearrangement of terms allows us to perform this integral
numerically much more quickly: 
\begin{equation}
\hat{K}_{j^{(*)}}=-\frac{i}{\hbar}\hat{U}_{0}(t)\int_{0}^{\infty}dt'\hat{U}_{0}^{-1}(t-t')\hat{H}'_{j^{(*)}}(t-t').\label{eq:Kj-def}
\end{equation}

\begin{figure}
\includegraphics[width=1\columnwidth]{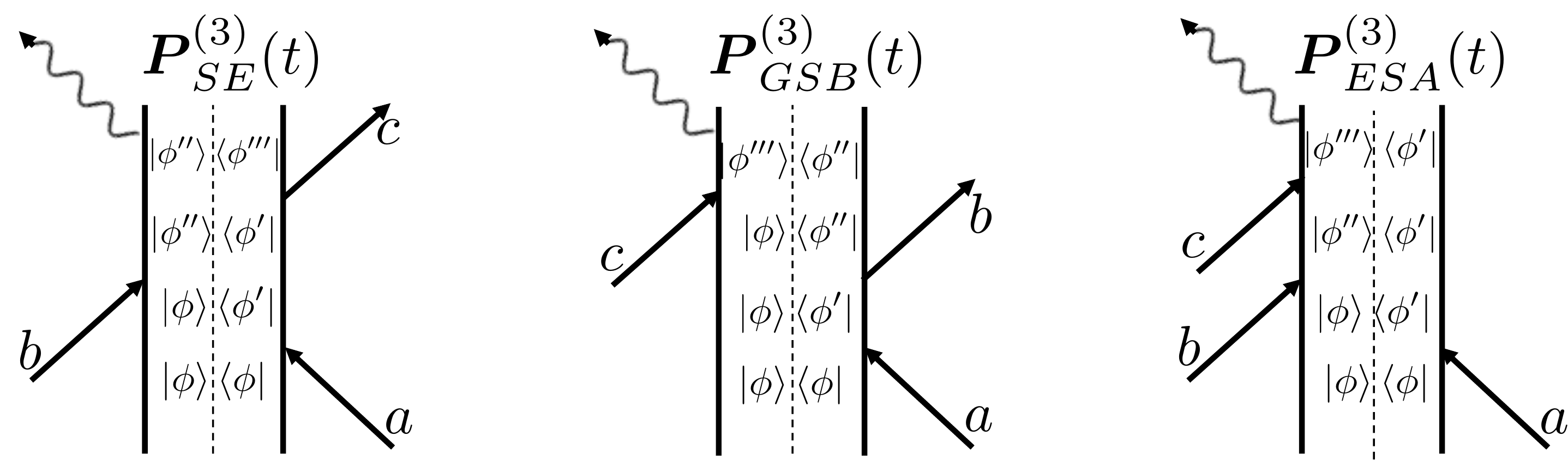}\caption{\label{fig:Three-Rephasing-Diagrams}From left to right, stimulated
emission (SE), ground-state bleach (GSB), and excited-state absorption
(ESA) double-sided Feynman diagrams for the rephasing 2D photon echo
signal. The dashed lines down the centers of each diagram emphasize
that the ket and bra evolve independently for closed systems. We calculate
the left and right sides of the diagram separately. These are the
only three diagrams that contribute to the rephasing signal when the
pulses are well-separated in time, though additional diagrams must
be considered when pulses overlap.}
\end{figure}

\begin{figure}
\includegraphics[width=1\columnwidth]{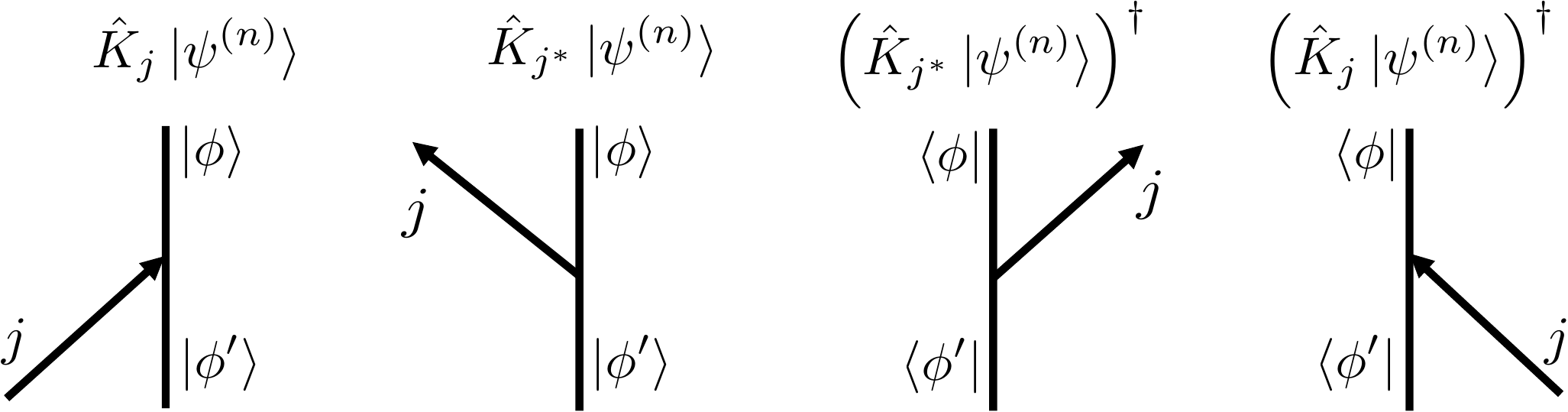}\caption{\label{fig:Kj-Pictures}Building blocks of Feynman diagrams for the
interaction with optical pulse $j$. The left two diagrams are for
kets and the right two are for bras. Time moves from bottom to top,
as the system begins in a linear combination of states $\protect\ket{\phi'}$,
and evolves to a linear combination of states $\protect\ket{\phi}$.
The arrow sloping up to the right is a contribution of the operator
$K_{j}$, and the arrow sloping up to the left is a contribution of
the operator $K_{j^{*}}$.}
\end{figure}

Most spectroscopic signals are not calculated using the full perturbative
wavefunction $\ket{\psi^{(n)}(t)}$, because only specific pathways
give nonzero contributions, which are visualized using Feynman diagrams,
as in Fig.~\ref{fig:Three-Rephasing-Diagrams}.\citep{Mukamel1999}
Each operator $\hat{K}_{j^{(*)}}$ represents a single interaction
arrow in a Feynman diagram, as in Fig.~\ref{fig:Kj-Pictures}. $\alg$
calculates the wavefunctions contributing to each diagram separately.
For example, as shown in Fig.~\ref{fig:Three-Rephasing-Diagrams},
the \diffbegin excited state absorption (ESA) contribution requires two wavefunctions
that we label $\ket{\psi_{bc}(t)}\equiv\hat{K}_{c}\hat{K}_{b}\ket{\psi^{(0)}}$
and $\ket{\psi_{a}(t)}\equiv\hat{K}_{a}\ket{\psi^{(0)}}$. \diffend

Following Eq.~\ref{eq:psi}, we write 
\[
\ket{\psi_{p}(t)}=\sum_{\phi'}e^{-i\omega_{\phi'}t}c_{\phi',p}(t)\ket{\phi'},
\]
where $p$ can be a multi-index, such as $ac^{*}$. Then we use Eqs.~\ref{eq:Hj}
and \ref{eq:Kj-def} to write \begin{widetext}
\begin{align}
\ket{\psi_{pj^{(*)}}(t)} & \equiv\hat{K}_{j^{(*)}}\ket{\psi_{p}(t)}\nonumber \\
 & =\frac{i}{\hbar}\hat{U}_{0}(t)\int_{0}^{\infty}dt'\hat{U}_{0}^{-1}(t-t')\sum_{\phi}\ket{\phi}\bra{\phi}\left(\hat{\cvec{\mu}}\cdot\cvec e_{j}^{(*)}\varepsilon_{j}^{(*)}(t-t')\right)\sum_{\phi'}e^{-i\omega_{\phi'}(t-t')}c_{\phi',p}(t-t')\ket{\phi'}\\
 & =\sum_{\phi}e^{-i\omega_{\phi}t}\ket{\phi}\frac{i}{\hbar}\int_{-\infty}^{\infty}dt'\theta(t')\underbrace{e^{i\omega_{\phi}(t-t')}\sum_{\phi'}\left(\cvec{\mu}_{\phi\phi'}\cdot\cvec e_{j}^{(*)}\varepsilon_{j}^{(*)}(t-t')\right)e^{-i\omega_{\phi'}(t-t')}c_{\phi',p}(t-t')}_{y_{\phi}(t-t')},\label{eq:Kj}
\end{align}
\end{widetext}where $\theta(t)$ is the unit step function. We rewrite
Eq.~\ref{eq:Kj} as
\begin{equation}
\ket{\psi_{pj^{(*)}}(t)}=\sum_{\phi}e^{-i\omega_{\phi}t}\ket{\phi}\underbrace{\frac{i}{\hbar}\left[\theta*y_{\phi}\right](t)}_{c_{\phi,pj^{(*)}}(t)},\label{eq:compact-Kj}
\end{equation}
where 
\[
\left[x*y\right](t)=\int_{-\infty}^{\infty}dt'x(t')y(t-t')
\]
is a convolution. Thus we arrive at a compact description of the coefficients
$c_{\phi,pj^{(*)}}(t)$. 
\begin{figure}
\includegraphics[width=1\columnwidth]{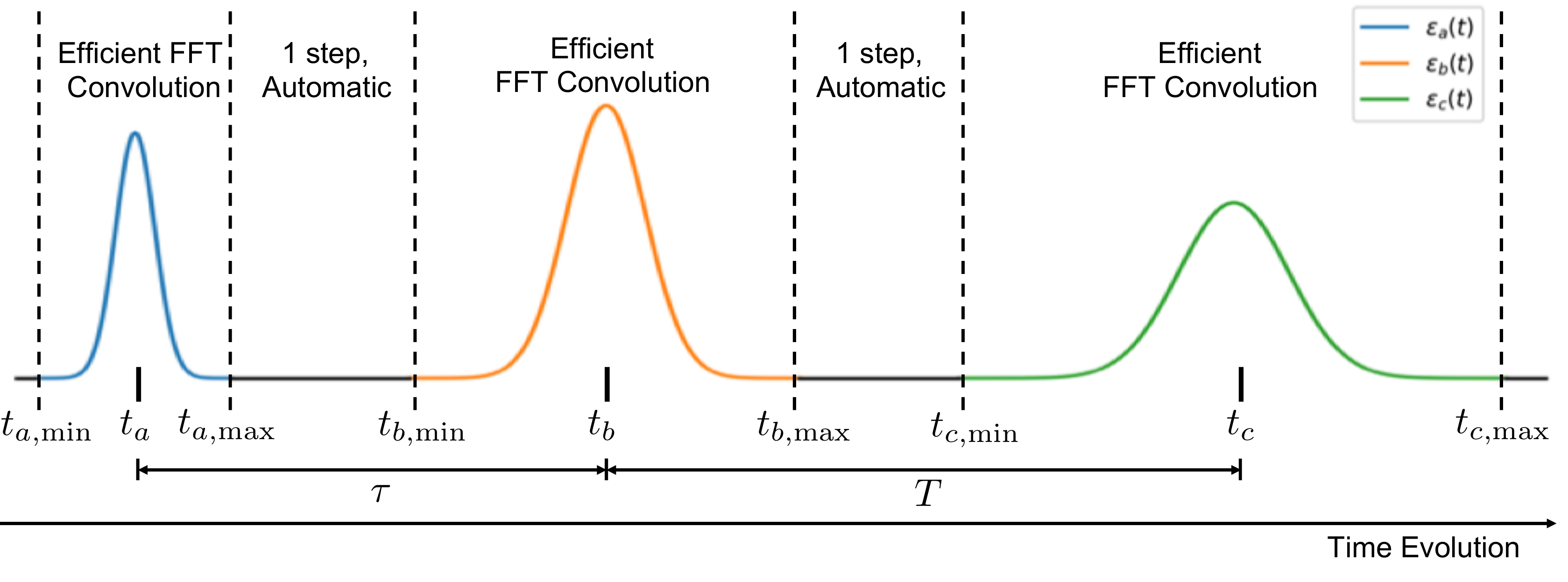}

\caption{\label{fig:PulseTrain}Sequence of three pulse envelopes, as would
be used in a rephasing 2DPE experiment, showing the time intervals
$(t_{j,\min},t_{j,\max})$ during which the pulses are non-negligible.
$\protect\alg$ only performs time propagations during these time
periods, using efficient FFT convolutions. At other times, the system
time evolution operator $\hat{U}_{0}$ gives exact time evolution.}
\end{figure}
If we make the physical assumption that the incident pulse is localized
in time, i.e., $\varepsilon_{i}(t)$ is negligible for $t<t_{j,\min}$
or $t>t_{j,\max}$, then it is also true that $y_{\phi,\alpha j^{(*)}}(t)$
is negligible when $t<t_{j,\min}$ or $t>t_{j,\max}$. This assumption
implies that 
\begin{equation}
c_{\phi,pj^{(*)}}(t)=\begin{cases}
0 & t<t_{j,\min}\\
r_{\phi}(t) & t_{j,\min}<t<t_{j,\max}\\
C_{\phi} & t>t_{j,\max}
\end{cases}.\label{eq:c_phi}
\end{equation}
Therefore, we need only calculate this convolution for $t_{j,\min}<t<t_{j,\max}$
(see Fig.~\ref{fig:PulseTrain}). Note that without the rearrangement
of Eq.~\ref{eq:Kj-def}, the convolution in Eq.~\ref{eq:compact-Kj}
would be more difficult to evaluate numerically because the functions
$c_{\phi,pj^{(*)}}(t)$ would include the phase factor $e^{-i\omega_{\phi}t}$
even after $t_{j,\text{max}}$ and thus not be constant. This trivial
definitional difference when working symbolically makes a large difference
when working numerically,\diffbegin and constitutes one of the primary novel
contributions of this work.

While Eq.~\ref{eq:Kj} is a rearrangement of the response function
formalism, and forms of it appear in References \onlinecite{Seibt2009,Yuen-Zhou2014,Smallwood2017,Perlik2017},
the integral in Eq.~\ref{eq:Kj} is usually incorporated into the
numerical propagation in direct propagation methods \citep{Engel1991,Meyer1999,Seibt2009,Renziehausen2009,Yuen-Zhou2014},
or is handled analytically \citep{Bell2015,Smallwood2017,Perlik2017,Do2017}.
By expressing this integral as a linear convolution over a small set
of points, we are able to include the effects of pulse shapes at a
modest cost, as shown in Appendix~\ref{sec:Computational-Cost-Analysis},
depending on the desired signal accuracy. \diffend

Physically, we only need to solve for the time dependence due to the
interaction while the pulse is non-zero. The rest of the time-dependence
is contained entirely in $\hat{H}_{0}$, and is therefore known exactly.
This realization drastically reduces the computational cost of $\alg$
compared to techniques that must use time-stepping for both the system
dynamics and the perturbation. We solve for the function $r_{\phi}(t)$
in Eq.~\ref{eq:c_phi} numerically using the convolution theorem
and FFT. \diffbegin With Gaussian pulse shapes of any duration, for convergence
of spectroscopic signals to a tolerance of 1\%, we find that 20-40
points is sufficient for calculating each convolution in Eq.~\ref{eq:compact-Kj}.
With such a small number of required points, the evaluation of the dipole
expectation value $\vec{P}(t)=<\psi|\hat{\mu}|\psi>$ is as expensive as the
propagation, as we detail in Appendix~\ref{sec:Computational-Cost-Analysis}.
\diffend

Translating the operators $K_{j^{(*)}}$ into computational algorithms
is straightforward. We discretize the time interval $(t_{j,\text{min}},t_{j,\text{max}})$
to have $M$ equally spaced points with spacing $dt$, which must
be small enough to resolve the pulses. Therefore, the wavefunctions
that give rise to the \diffbegin ESA \diffend signal are calculated in three steps: \diffbegin
\begin{enumerate}
\item Calculate $\ket{\psi_{a}(t)}=\hat{K}_{a}\ket{\psi^{(0)}}$,
\item Calculate $\ket{\psi_{b}(t)}=\hat{K}_{b}\ket{\psi^{(0)}}$ for fixed
$\tau$,
\item Calculate $\ket{\psi_{bc}(t)}=\hat{K}_{c}\ket{\psi_{b}(t)}$ for fixed
$T$.
\end{enumerate} \diffend
The polarization field is then \diffbegin
\begin{equation}
\cvec P_{ESA}^{(3)}(t)|_{\tau,T}=\bra{\psi_{a}(t)}\hat{\cvec{\mu}}\ket{\psi_{bc}(t)}|_{\tau,T}.\label{eq:P_of_t_ESA}
\end{equation} \diffend
This process is repeated for all $\tau$ and $T$ of interest.

We demonstrate how this conceptual procedure is implemented using
the syntax of $\alg$.
\diffbegin
\lstset{language=Python,basicstyle=\ttfamily}
\begin{lstlisting}[basicstyle=\small]
psi_a = uf2.up(uf2.psi0,pulse_number = 0) 
psi_b = uf2.up(uf2.psi0,pulse_number = 1) 
psi_bc = uf2.up(psi_b,pulse_number = 2)
P_ESA = uf2.dipole_expectation(psi_a,psi_bc)
S_ESA = uf2.polarization_to_signal(P_ESA)
\end{lstlisting} \diffend
In the syntax of $\alg$, pulse $a$ is pulse 0, pulse $b$ is pulse
1, etc., and the \lstinline{up} and \lstinline{down} methods implement
$K_{j}$ and $K_{j^{*}}$, respectively. This snippet is adapted from
the Jupyter notebook Example.ipynb included with the source code.

One of the strengths of $\alg$ is that the user interacts with it
by translating Feynman diagrams into nested calls to the $\hat{K}_{i^{(*)}}$
functions. The goal of this algorithm is to make calculating spectra
as easy as writing down Feynman diagrams. We have used $3^{\text{rd}}$-order
diagrams for the derivation because they are the most familiar, but
calculating arbitrary order diagrams is just as straightforward. $\alg$
is therefore easy to use and avoids the necessity of first writing
out lengthy symbolic expressions for each diagram and translating
each expression into code. $\alg$ can also be used in the impulsive
limit to calculate response functions, allowing this ease of use to
be applied to the study of response functions as well.

\subsection{Optical dephasing\label{subsec:Including-Optical-Dephasing}}

For closed systems, $\cvec P^{(n)}(t)$ oscillates for all time after
the $n$ optical interactions. This undamped oscillation is unphysical.
In order to include optical dephasing in general, we would need to
solve for the third-order density matrix, $\rho^{(3)}(t)$, instead
of the wavepackets. \diffbegin In order to include optical dephasing in our polarization
signals, we convolve $\tilde{\cvec P}^{(n)}(\omega)$ with a lineshape
function, as in Ref.~\onlinecite{Seibt2009}. Whereas that work used
a Gaussian shape to model inhomogeneous broadening, in the examples
below we use a Lorentzian lineshape to model homogeneous broadening
due to an exponential dephasing. \diffend

\subsection{Useful standard approximations\label{subsec:RWA-PM}}

The derivations in Section~\ref{subsec:Derivation} do not require
the rotating wave approximation (RWA) or phase matching. However,
these approximations significantly reduce the cost of the calculations,
as in all perturbative spectroscopies.\citep{Cho2009,PeterHamm2011,Mukamel1999} \diffbegin
In the RWA, only one of either the rotating or counter-rotating terms
of Eq.~\ref{eq:Hdecomposition} contributes to an interaction.\citep{PeterHamm2011}
\footnote{The rotating term excites a ket and de-excites a bra. The counter-rotating
term de-excites a ket and excites a bra.} \diffend Since $\ket{\psi^{(n)}(t)}\propto\left(\hat{H}'(t)\right)^{n}$,
the $n^{\text{th}}$-order wavefunction is composed of $(2L)^{n}$
terms. The standard RWA and phase-matching conditions reduce the number
of terms relevant to spectroscopic signals. The RWA is valid in the
limit that the pulse durations are long compared to the optical carrier
frequencies, which are roughly degenerate with the optical energy
gap of $\hat{H}_{0}$. If the material system is dispersed over a
volume much larger than the wavelength of the light then the phase-matching
condition ensures that signal fields will be produced only in directions
corresponding to sums and differences of wavevectors of the optical
pulses \citep{Mukamel1999}. For example Fig.~\ref{fig:Three-Rephasing-Diagrams}
shows the signals produced in the rephasing direction ($-\cvec k_{a}+\cvec k_{b}+\cvec k_{c}$)
of 2D photon echo (2DPE) spectroscopy. 

In addition to reducing the number of relevant diagrams, the RWA speeds
up calculations because we do not need to keep track of the optical
carrier frequency. This advantage is significant because $\alg$ performs
calculations using $M$ time points with spacing $dt$ from $t_{j,\min}$
to $t_{j,\max}$. In the RWA, we set the optical carrier frequency
to 0, and therefore $dt$ only needs to resolve the pulse envelope
and not the carrier frequency. 

\subsection{Efficiency improvements\label{subsec:Efficiency-Improvements}}

We increase the efficiency of $\alg$ by decreasing the required dimensionality
of the Hilbert space from the full size, $N_{\text{total}}$, to a \diffbegin
compressed subspace of dimension $N_{c}$, \diffend over which the sums in
Eq.~\ref{eq:Kj} run. \diffbegin We perform the reduction from $N_{\text{total}}$
to $N_{c}$ before running any propagations \diffend by determining which elements
of the dipole operator $\hat{\cvec{\mu}}$ will not contribute appreciably
to the calculation and ignoring them. We can prune the required set
of states in two ways. First, many systems have elements $\cvec{\mu}_{\phi'\phi}\approx0$.
Second, we determine which energetic transitions will not be allowed
by the electric field shape.

We use Bessel's inequality to define which states $\ket{\phi'}$ are
important in order to resolve $\cvec{\mu}_{\phi'\phi}$ accurately.
If we begin in the eigenstate $\text{\ensuremath{\ket{\phi}}}$, one
interaction with the dipole operator (as occurs in each application
of $K_{j^{(*)}}$) yields the state 
\begin{align*}
\ket{\Phi} & =\hat{\cvec{\mu}}\ket{\phi}=\sum_{\phi'=1}^{N_{\text{total}}}\cvec{\mu}_{\phi'\phi}\ket{\phi'}.
\end{align*}
We seek to restrict the sum to the smallest number of terms without
significantly altering the norm of this state, which implies that
we have captured all of the physically relevant states. We write the
norm of $\ket{\Phi}$ as
\[
\braket{\Phi}{\Phi}=\sum_{\phi'=1}^{N_{\text{total}}}\abs{\cvec{\mu}_{\phi'\phi}}^{2}=\bra{\phi}\hat{\cvec{\mu}}^{2}\ket{\phi}=\left(\cvec{\mu}^{2}\right)_{\phi\phi}.
\]
We find the smallest set of states, \diffbegin $N_{c}$, \diffend such that 
\begin{equation}
\frac{\left(\cvec{\mu}^{2}\right)_{\phi\phi}-\sum^{'}\abs{\cvec{\mu}_{\phi\phi'}}^{2}}{\left(\cvec{\mu}^{2}\right)_{\phi\phi}}>1-\epsilon,\label{eq:pruning-metric}
\end{equation}
for small $\epsilon$, where $\sum'$ represents the restricted sum
over the required \diffbegin $N_{c}$ \diffend states. We perform this analysis for all
required states $\ket{\phi}$.\textbf{ }

We make this concept concrete by giving an example using the non-rephasing
ground state bleach signal, which is not included in Fig.~\ref{fig:Three-Rephasing-Diagrams}.
The polarization field produced by that diagram is $\cvec P_{GSB,NR}^{(3)}(t)=\bra{\psi^{(0)}}\cvec{\mu}\ket{\psi_{ab^{*}c}(t)}$.
If we begin, say, in the state $\ket{\psi^{(0)}}=\ket 1$, the lowest
energy eigenstate of $\hat{H}_{0}$, then we must determine the smallest
number of states, \diffbegin $N_{c}^{(1)}$, \diffend that satisfy Eq.~\ref{eq:pruning-metric}
for $\phi=1$. Then we know before calculating Eq.~\ref{eq:Kj} that
$\ket{\psi_{a}(t)}=\sum c_{\phi,a}(t)\ket{\phi}$ will be composed
of $N_{c}^{(1)}$ terms. To find the states required to describe $\ket{\psi_{ab^{*}}}$,
we use Eq.~\ref{eq:pruning-metric} for each of the $N_{c}^{(1)}$
states needed for $\ket{\psi_{a}(t)}$, which gives a new set of $N_{c}^{(2)}$
states, where the \diffbegin superscript \diffend indicates the number of times $\hat{\mu}$
has been applied to the initial state.

The next obvious step is to use Eq.~\ref{eq:pruning-metric} for
each of those $N_{c}^{(2)}$ states, but that step is actually unnecessary.
Such an analysis would allow us to determine the $N_{c}^{(3)}$ states
required to resolve $\ket{\psi_{ab^{*}c}(t)}$. However, the final
signal depends upon $\bra{\psi^{(0)}}\cvec{\mu}\ket{\psi_{ab^{*}c}(t)}$,
and therefore the only components of $\ket{\psi_{ab^{*}c}(t)}$ that
matter spectroscopically are those that overlap with $\cvec{\mu}\ket{\psi^{(0)}}$.
These are precisely the $N_{c}^{(1)}$ states we determined in the
first step of this process. Therefore we only need the same $N_{c}^{(1)}$
states when calculating $\ket{\psi_{ab^{*}c}(t)}$.

For many systems relevant to optical spectroscopy, there are well-separated
manifolds of states with 0, 1, 2, etc.\ electronic excitations. When
there are well-separated manifolds, the RWA allows a further reduction
to the relevant size $N_{c}$. In the same non-rephasing GSB example,
$\ket{\psi_{a}(t)}$ is in the singly excited manifold (SEM), and
$\hat{\mu}\ket{\psi_{a}(t)}$ has components in both the ground state
manifold (GSM) and the doubly excited manifold (DEM). In the RWA,
however, $\ket{\psi_{ab^{*}}(t)}$ is only in the GSM, so $N_{c}^{(2)}$
can be divided into its GSM and DEM portions, with only the GSM portion,
$N_{c,GSM}$, required for this diagram. The DEM portion of $N_{c}^{(2)}=N_{c,DEM}$
is required for the ESA diagram.

In addition, we can use basic knowledge of the shape of the electric
field to further restrict $N_{c}$. $\alg$ performs discrete convolutions
for times $t$ with spacing $dt$ (see end of Section~\ref{subsec:RWA-PM}).
The spacing $dt$ is chosen in order to resolve the shape of the electric
field and implies a frequency range in which $\tilde{E}(\omega)$
is non-zero. The maximum frequency resolved is $\frac{\pi}{dt}$.
Each element $\cvec{\mu}_{\phi'\phi}$ has an associated frequency
difference $\omega_{\phi'}-\omega_{\phi}$. If $\abs{\omega_{\phi'}-\omega_{\phi}}\leq\frac{\pi}{dt}$,
then the transition is energetically allowed by the pulse. If $\abs{\omega_{\phi'}-\omega_{\phi}}>\frac{\pi}{dt}$,
the transition is not allowed, and we set $\cvec{\mu}_{\phi'\phi}=0$
for all those energetically inaccessible transitions, further decreasing
the relevant size $N_{c}$.\footnote{This procedure is not only helpful for speeding up calculations, but
actually necessary for accurate calculations. If included, any frequency
differences $\abs{\omega_{\phi'}-\omega_{\phi}}>\frac{\pi}{dt}$ would
cause spurious signals to appear in the signal due to aliasing.} For long-duration pulses, this reduction is highly significant and
allows treatment of long propagation times without increased cost.

\diffbegin In practice, to obtain the eigenstates of $\hat{H}_{0}$, it generally
must first be truncated from dimension $N_{\text{total}}$ to a finite
dimension $N_{f}$, which must be chosen to be large enough that the
$N_{c}$ required eigenstates are resolved sufficiently accurately.
Obtaining the $N_{c}$ eigenstates requires diagonalizing all or part
of the truncated Hamiltonian of size $N_{f}$. Full diagonalization
is well-known to scale as $N_{f}^{3}$. Since $N_{f}$ is sparse,
iterative methods can be used to obtain the necessary subset $N_{c}$,
iterative methods can give better scalings in some cases. We will
present our method for diagonalizing vibronic systems, including anharmonicities
and varying vibrational frequencies, in a separate manuscript. \diffend The
reduction from dimension $N_{f}$ to $N_{c}$ not only makes $\alg$
more efficient but also dramatically expands the range of systems
for which $\alg$ is tractable. \diffbegin In Section~\ref{sec:Example} we
calculate TA spectra for a vibronic system, which formally has $N_{\text{total}}=\infty$.
In practice we require $N_{f,DEM}=N_{f,GSM}=15$ and $N_{f,SEM}=30$.
With $\epsilon=0.001$ the spectra converge to within $1\%$. Using
the procedure outlined here, we determine that $N_{c,SEM}=8$ and
$N_{c,DEM}=N_{c,GSM}=10$. $N_{c,SEM}$ would be 35 in order to correctly
resolve $\ket{\psi_{ab^{*}c}(t)}$, but we only need to use $N_{c,SEM}=8$
of those states to accurately reproduce the TA spectra.

\section{Computational Cost\label{sec:Computational-Advantage}}

In this section we discuss in brief the computational cost of $\alg$
and compare it to two alternative methods of obtaining nonlinear spectra:
direct propagation (DP) of the Schrodinger equation without diagonalization
of $\hat{H}_{0}$ and a Fourier-space pseudospectral method. These
alternative methods are commonly used in predictions of nonlinear
spectroscopies and each has a domain where it is the most efficient
method.

In contrast to $\alg$, DP methods propagate the perturbative wavefunction
using the Schrodinger equation in the form 
\begin{equation}
\frac{d\ket{\psi^{(n)}(t)}}{dt}=-\frac{i}{\hbar}\hat{H}_{0}\ket{\psi^{(n)}(t)}-\frac{i}{\hbar}\hat{H}'_{j}(t)\ket{\psi^{(n-1)}(t)}.\label{eq:TDSEdiff}
\end{equation}
$\alg$ performs the same integration in the eigenbasis, where Eq.~\ref{eq:TDSE}
permits rapid evaluation. There are a number of methods of directly
integrating Eq.~\ref{eq:TDSEdiff}, and to give a sense of where
$\alg$ is most effective, we compare here to an adaptive step-size
4-5$^{\text{th}}$ order Runge-Kutta solver for the $\hat{H}_{0}$
term and an Euler method for the perturbation, which is described
in Section~\ref{sec:RK45Implementation} and which we call the RKE
method. We believe the general scaling trends of these results will
be similar for other DP methods, whether they be Adams-Bashforth \citep{Johansson2012},
Burlisch-Stoer \citep{Tsivlin2006}, or others. While the above derivation
is general, we perform these comparisons for sets of vibronic systems,
which we introduce in Section~\ref{subsec:Properties-of-Vibronic}.
In all cases, we use the RWA to increase step sizes as much as possible.

The cost of obtaining all eigenvalues and eigenvectors of a matrix
of dimension $N_{f}$ scales as $N_{f}^{3}$ for large $N_{f}$. If
$N_{f}$ is sparse, efficient iterative algorithms exist to extract
the subspace $N_{c}$, which can scale more favorably. In contrast,
direct propagation scales as $N_{f}$ in the case that $\hat{H}_{0}$
is sparse and $N_{f}^{2}$ if $\hat{H}_{0}$ is not sparse. Therefore
it is clear that for sufficiently large $N_{f}$ the DP methods will
eventually be the more efficient option. We find, however, that $N_{f}$
needs to be quite large before this crossover occurs. Defining this
break-even size is somewhat difficult, as diagonalization only needs
to be performed once, and the timing comparison depends upon how many
times the eigensystem is to be reused. In Section \ref{sec:Example},
we compute isotropically averaged TA spectra of a model system for
hundreds of different dipole moments and 3 different electric field
pulse shapes, which requires thousands of diagrams to be calculated,
all with the same $\hat{H}_{0}$. Even if the diagonalization time
were greater than the cost of obtaining a single spectrum, that diagonalization
cost may be negligible compared to the cost of producing all of the
spectral predictions. It is also therefore important to compare the
cost of $\alg$ separately from the cost of diagonalization. The cost
of propagation with $\alg$ is also asymptotically worse than DP methods,
but we show in Sec.~\ref{subsec:UF2-vs-RKE} that $\alg$ remains
faster for vibronic systems up to size $\sim10^{4}$. For frequently
studied systems with dimension less than 100, as in Sec.~\ref{sec:Example},
$\alg$ can be over 100 times faster than direct propagation methods.

\subsection{Properties of Vibronic Systems\label{subsec:Properties-of-Vibronic}}

The relative costs of propagation methods depend on the structure
of the Hamiltonian, and we use vibronic systems for our examples.
Vibronic systems consist of two or more coupled electronic states,
each of which is coupled to one or more vibrational degrees of freedom,
which are often harmonic. There is no known general solution to the
time-independent Schrodinger equation for such systems. A common choice
of basis for DP methods is the number basis of the harmonic oscillator
of each vibrational mode, and in this basis, $H_{0}$ is highly sparse
\citep{Domcke2007}. This basis is infinite and therefore must be
truncated to some finite size, which we call $N_{f}$.

For the purposes of cost comparisons, we consider vibronic systems
with optically separated manifolds. We focus in particular on $s$
coupled two-level systems (TLS), each with an electronic ground state,
$|g\rangle$, and a single optical excitation $\ket s$, described
by the Hamiltonian
\begin{align*}
\hat{H}_{e} & =\ket g\bra g+\sum_{i=1}^{s}E_{i}\ket i\bra i+\sum_{i\neq j}J_{ij}\ket i\bra j,
\end{align*}
where $E_{i}$ is the site energy and $J_{ij}$ is a Hermitian matrix
of couplings. Vibrations are described by the Hamiltonian
\[
\hat{H}_{ph}=\frac{1}{2}\left(\sum_{\alpha=1}^{k}\hat{p}_{\alpha}^{2}+\omega_{\alpha}^{2}\hat{q}_{\alpha}^{2}\right),
\]
where $k$ is the number of independent vibrations, and $\hat{p}_{\alpha}$
and $\hat{q}_{\alpha}$ are the generalized momentum and coordinate
operators for each vibration, with frequencies $\omega_{\alpha}$.
Standard linear coupling of the position of each oscillator to its
site excitation gives 
\[
\hat{H}_{e-ph}=\sum_{i=1}^{s}\sum_{\alpha=1}^{k}\omega_{\alpha}^{2}d_{\alpha,i}q_{\alpha}\vert i\rangle\langle i\vert,
\]
where $d_{\alpha,i}$ are the coupling strengths, corresponding to
Huang-Rhys factors
\[
S_{\alpha,i}=\frac{1}{2}\omega_{\alpha}d_{\alpha,i}^{2}.
\]
The total system Hamiltonian is 
\begin{equation}
\hat{H}_{0}=\hat{H}_{e}+\hat{H}_{ph}+\hat{H}_{e-ph}.\label{eq:nmer}
\end{equation}
This Hamiltonian is block diagonal, and in third-order spectroscopies
there are three relevant manifolds: the GSM, SEM, and DEM as described
in Section~\ref{subsec:Efficiency-Improvements}, with higher manifolds
becoming relevant in higher-order spectroscopies. Only the perturbation,
$\hat{H}'$, mixes these manifolds. Propagation within each manifold
can be handled independently, and each manifold can be truncated with
dimension $N_{f,X}$, where $X$ can be GSM, SEM, DEM, .... If we
consider only the first three manifolds, then $N_{f}=N_{f,GSM}+N_{f,SEM}+N_{f,DEM}$.

The relevant dimension $N_{f,SEM}$ can be dramatically smaller than
$N_{f,GSM}$ and $N_{f,DEM}$. The reason for this smaller size can
be illustrated with the case $s=k=1$. In this case, the vibrational
system can be diagonalized when the electron is in state $\ket g$,
giving a vibrational basis $\ket{g,n}$, where $n$ is the vibrational
quantum number; it can be rediagonalized when the electron is in the
excited state $\ket e$ to give a vibrational basis $\ket{e,n}$,
and the union of the two bases is a basis for the full system. For
the first pulse interaction, with small values of $d$, the ground
vibrational state $\ket{g,0}$ only couples to the first few vibrational
levels of the excited state via the dipole interaction (e.g., $\ket{e,0},\ket{e,1},\ket{e,2}$),
due to the rapidly decaying Franck-Condon overlaps. With the second
pulse interaction, these three levels in turn couple to the bottom
5 or 6 ground state levels. Each optical transition couples more vibrational
levels. However, as described in Section~\ref{subsec:Efficiency-Improvements},
4-wave mixing signals only require correctly resolving amplitudes
in the basis states required for the first- and second-order wavepackets.
In this case, the required numbers of states in the manifolds are
different, so $N_{f,SEM}<N_{f,GSM}$ and $N_{c,SEM}<N_{c,GSM}$ .
This result remains true for vibronic systems with $s>1$ and $k>1$,
and similar arguments lead to the conclusion that $N_{c,SEM}<N_{c,DEM}$.

\subsection{Comparison to Direct Propagation\label{subsec:UF2-vs-RKE}}

\begin{table}
\caption{\label{tab:Summary-of-symbols} \diffbegin Summary of symbols used to describe
computational cost of a 2DPE rephasing signal. $M$ and $M_{t}$ are
convergence parameters. \diffend}

\begin{tabular}{cc}
 & \tabularnewline
\hline 
\hline 
Symbol & Definition\tabularnewline
\hline 
$N_{f,X}$ & Full size of Hilbert space of manifold $X$\tabularnewline
$N_{c,X}$ & Compressed size of Hilbert space of manifold $X$\tabularnewline
$M_{t}$ & Number of time points to resolve $\cvec P(t)$ in $\alg$\tabularnewline
$M_{RK}$ & Number of time points to resolve $\cvec P(t)$ in RKE\tabularnewline
$\alpha$ & Cost of complex floating point multiplication\tabularnewline
$m_{\tau}$ & Number of desired values of $\tau=t_{b}-t_{a}$\tabularnewline
$m_{T}$ & Number of desired values of $T=t_{c}-t_{b}$\tabularnewline
\hline 
\hline 
 & \tabularnewline
\end{tabular}
\end{table}
\begin{table}
\caption{\label{tab:Summary-of-symbols-2} \diffbegin Summary of symbols used to describe
computational cost of a 2DPE rephasing signal. Particular values are
stated for the range of $k=2$-$8$ vibrations, corresponding to the
systems studied in Figure~\ref{fig:Cost-Ratio}. \diffend}

\begin{tabular}{ccc}
 &  & \tabularnewline
\hline 
\hline 
Symbol & Definition & Range\tabularnewline
\hline 
$l_{SEM}$ & $N_{c,SEM}/N_{f,DEM}$ & 2-60\tabularnewline
$l_{DEM}$ & $N_{c,DEM}/N_{f,DEM}$ & $\approx2$\tabularnewline
$M_{RK}/M_{t}$ & Ratio of field-free step sizes & 2-10\tabularnewline
$r$ & Number of nonzero entries per row of $\hat{H}_{0}$ & $2k+1$\tabularnewline
$q$ & Sparse matrix overhead & $\approx2$\tabularnewline
\hline 
 &  & \tabularnewline
\end{tabular}
\end{table}

The most difficult part of calculating any $n$-wave mixing signal
is the calculations that correspond to the last two arrows of a Feynman
diagram. In Fig.~\ref{fig:Three-Rephasing-Diagrams}, for example,
these are the interaction with pulse $c$ and the emission of the
polarization field. These steps are the most costly because they must
be repeated for all $m_{\tau}$ desired values of delays between pulses
$a$ and $b$ and all $m_{T}$ desired values of delays between pulses
$b$ and $c$.

In Appendix A, we show in Eq.~\ref{eq:UF2-ESA-cost} that the cost
of calculating third-order signals using $\alg$ scales asymptotically
as 
\begin{equation}
C_{\alg}=m_{\tau}m_{T}\alpha N_{c,SEM}N_{c,DEM}M_{t},\label{eq:C_UF2}
\end{equation}
where $\alpha$ is the cost of multiplying two complex floating-point
numbers, and symbols used in this section are defined in Table~\ref{tab:Summary-of-symbols}.
In Appendix B, we introduce the RKE method and show in Eq.~\ref{eq:RKE-ESA-cost}
that the same calculation using RKE scales asymptotically as 
\begin{equation}
C_{RKE}=6m_{\tau}m_{T}\alpha qrN_{f,DEM}M_{RK},\label{eq:C_RK}
\end{equation}
where $q$, $r$, and $M_{RK}$ are defined in Table.~\ref{tab:Summary-of-symbols-2}.

We now estimate the break-even system size where RKE becomes less
expensive than $\alg$ and compare to timings with our sample systems.
In each of our comparisons, the nonzero $d_{\alpha,i}$ are identical,
and the $\omega_{\alpha}$ are within 10\% of one another. We study
systems with $s$ ranging from 2 to 8, $k$ ranging from 2 to 8, and
$k/s$ ranging from 1 to 4. We also study $S_{\alpha,i}$ ranging
from 0.02 to 4.5, with $S_{\alpha,i}$ equal for all modes in a given
system. We use a homogeneous linewidth of $0.1\omega_{0}^{-1}$, where
$\omega_{0}$ is the smallest value of $\omega_{\alpha}$. We use
identical pump and probe pulses with a Gaussian profile centered on
the transition from the lowest energy GSM eigenstate to the lowest
energy SEM eigenstate. Both pulses have a Gaussian time-domain standard
deviation of $\sigma=\omega_{0}^{-1}$.

To compare Eqs.~\ref{eq:C_UF2}-\ref{eq:C_RK}, we must relate $N_{c,X}$
to $N_{f,X}$ and $M_{t}$ to $M_{RK}$. In order to correctly determine
the $N_{c,X}$ eigenvalues and eigenvectors in manifold $X$, $N_{f,X}$
must be made large enough. For third-order signals to converge to
better than 1\%, we find that $N_{c,X}\lesssim N_{f,X}/2$ in the
systems we have studied. Stricter convergence requirements tend to
leave $N_{c,X}$ unchanged while increasing the required $N_{n,X}$.
As a wavefunction propagates with a DP method, it obtains weights
in ever higher vibrational states and thus requires $N_{f,X}$ to
be sufficiently large to ensure accurate calculations; this requirement
is similar to the requirement that $N_{f,X}$ must be large enough
to resolve the relevant $N_{c,X}$ eigenvalues of the system, and
we assume that, given the accuracy of the predicted spectrum that
is desired, the required $N_{f,X}$ are approximately equal for both
$\alg$ and DP methods. An important factor in determining $N_{f}$
is the longest propagation time required. The required $N_{f}$ is
smaller for a TA spectrum that includes delay times out to $2T_{\alpha}$
than it is for a TA spectrum that includes delay times out to $15T_{\alpha}$,
where $T_{\alpha}=2\pi/\omega_{\alpha}$ is the vibrational period.

\begin{figure}
\includegraphics[width=1\columnwidth]{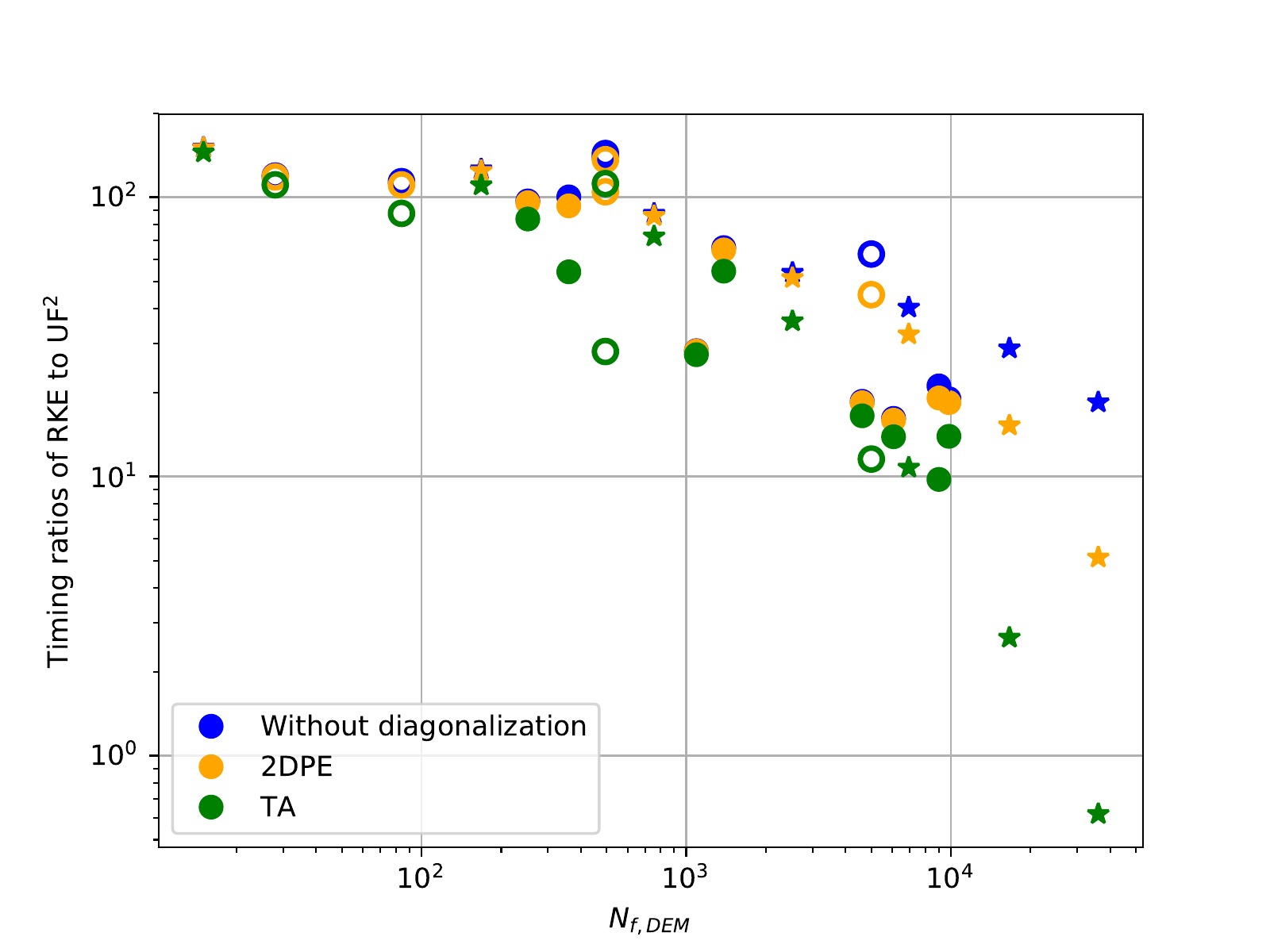}\caption{\label{fig:Cost-Ratio} \diffbegin Ratio of runtimes of RKE to $\protect\alg$
for an array of vibronic systems, as described in the text. We include
the diagonalization time when we calculate a single TA spectrum with
90 delay times, and a single 2DPE with 50 values of $\tau$ and 20
values of $T$. In most cases, the cost of diagonalization is negligible.
Stars indicate systems with $s=k$ ranging from 2 to 8 with $S=0.02$.
Open circles are systems with $s=2$ and $k$ ranging from 2 to 8,
with $S$ ranging from 0.02 to 0.08. Filled circles are systems with
$s=3$ and $k=3$ or $k=6$, with $S$ ranging from 0.02 to 4.5. \diffend}
\end{figure}

For the numerical comparisons in Fig.~\ref{fig:Cost-Ratio}, we use
both $\alg$ and RKE to calculate spectra for $m_{T}=90$ delay times between
0 to $89\omega_{\alpha}^{-1}$, which is half the number used for all
calculations in
Section~\ref{sec:Example}. Both $\alg$ and RKE scale linearly with
the number of delay times, so the number of delay times is important
only for the ratio including the cost of diagonalization. For the
cost of calculating a 2DPE, we scale the cost per delay time by 1000,
corresponding to $m_{\tau}=50$ and $m_{T}=20$.

We write $N_{c,SEM}=\frac{N_{f,DEM}}{l_{SEM}}$ and $N_{c,DEM}=\frac{N_{f,DEM}}{l_{DEM}}$,
where $l_{SEM}$ ranges from $2$ for $k=s=2$ and $S=0.02$, to $60$
for $k=s=8$ and $S=0.02$. For all the cases studied here, $l_{DEM}\approx2$.

Ignoring the small effect from the $J_{ij}$ couplings, given $k$
vibrational modes, $r=2k+1$. Generally $M_{RK}>M_{t}$, since $\alg$
is limited only by the pulse, whereas RKE is limited both by the pulse
and the system. In particular, the RKE step size, and therefore $M_{RK}$,
is limited by the largest eigenvalues included in the truncated Hamiltonian
of size $H_{f}$. We have found in the sparse matrix implementation
of scipy running on a MacBook Pro, 2.3 GHz Intel Core i5, that $q\approx2$.
We have also found for systems ranging from $s=k=2$ to $s=k=8$ that
$M_{RK}/M_{t}$ ranges from 2 to 10. The ratio of the costs of the
two methods is then
\begin{equation}
\frac{C_{RK}}{C_{\alg}}=6qrl_{SEM}l_{DEM}\frac{M_{RK}}{M_{t}}\frac{1}{N_{f}},\label{eq:Cost-Ratio}
\end{equation}
which predicts that $\alg$ will outperform the RKE solver until the
system Hamiltonian reaches a size in the range of $10^{3}-10^{5}$.
This estimate is rough, and depends upon the details of each system.

In Fig.~\ref{fig:Cost-Ratio} we plot the ratio of observed runtimes
for RKE to $\alg$ for a range of different vibronic systems. Simulations
were run on an Intel Xeon E5-2640 v4 CPU with a 2.40GHz clock speed.
We compare $\alg$ and RKE by calculating spectra with better than
1\% convergence for $\alg$, and better than 5\% convergence for RKE.
We hold RKE to a lower standard because our implementation requires
the same time step while the pulse is on and while the pulse is off.
We have found that for RKE to reach $1\%$ convergence, $dt$ must
be so small that it fails to take advantage of the adaptive RKE step
size. We therefore run tests in a regime where RKE can take advantage
of its adaptive step size, in order to give a fair comparison. To
achieve $1\%$ convergence in its current form, RKE would take about
4-5 times longer to run for all of the cases in Fig.~\ref{fig:Cost-Ratio},
and we are not sure that extra time can truly be eliminated, so we
are giving the RKE method an advantage in Fig.~\ref{fig:Cost-Ratio}.

The results in Figure \ref{fig:Cost-Ratio} indicate that $\alg$
is 150 times faster than RKE for systems with $N_{f,DEM}<100$ and
is competitive until $N_{f,DEM}$ is over $10^{4}$, with the exact
break-even point depending on how many different times the eigenvectors
can be reused. For all but the largest systems studied, the diagonalization
cost is not a significant contribution to runtimes. Including rotational
or thermal averaging, or variations in $\hat{\mu}$ would further
reduce the importance of the diagonalization costs. Systems with required
dimensions over $10^{5}$ would benefit from the RKE method.

We note that the cost of the current implementation of $\alg$ is
determined by resolving the final polarization with $M_{t}$ points
having the same step size $dt$ as needed to resolve the last pulse
interaction. There is no fundamental reason that the polarization
must be resolved with the same time spacing as the last optical pulse,
and we anticipate that future developments could reduce the $\alg$
runtimes by a further factor of 5, until the wavefunction propagation
and final polarization evaluation have similar cost.

\subsection{Comparison to split-operator pseudospectral methods\label{subsec:Split-operator}}

Another large class of methods useful for numerical prediction of
nonlinear spectroscopies is pseudospectral methods, which use real-
or Fourier-space representations of nuclear wavepackets, rather than
explicit eigenstates or a basis of vibrational quanta \citep{Domcke2007,Feit1982,Kosloff1983}.
The widely used split-operator pseudospectral method (SOP) uses a
plane-wave basis and an associated evenly-spaced real-space basis,
with the efficient FFT to move between them. Briefly, SOP performs
part of the time evolution due to $\hat{H}_{0}$ in position space
and part in momentum space, with FFT operations between. The perturbative
interaction with the pulses is also handled numerically, using a Euler
method similar to that in Appendix \ref{sec:RK45Implementation}.
The time step $dt$ must be short enough to resolve both the system
and perturbative dynamics. It is not trivial to compare $\alg$ directly
to SOP methods, as the convergence parameters (e.g., $N_{c}$ for
$\alg$ and the grid-spacing in SOP) are not directly comparable.
We compare $\alg$ to the SOP implementation detailed in Ref.~\onlinecite{Yuen-Zhou2014}.

For problems where a small number of eigenstates are required to describe
the system dynamics, the eigenbasis used by $\alg$ is considerably
more efficient than a real/Fourier-space representation, which can
still need many discretization points to resolve the wavepackets.
In such problems, which include most of the class of problems described
by Eq.~\ref{eq:nmer}, we expect $\alg$ to greatly outperform SOP
methods. Additionally, since $\alg$ handles the evolution due to
$\hat{H}_{0}$ exactly, $\alg$ can evolve the wavefunctions forward
in the absence of the pulses for near zero cost. In addition, the
time step $dt$ used by $\alg$ to evaluate the convolutions during
the pulses is determined solely by the pulse shapes and need not be
small enough to accurately resolve the dynamics in $\hat{H}_{0}$.
For the cases shown in Section~\ref{sec:Example}, which involve
harmonic vibrational modes, $\alg$ calculates spectra at least $10^{5}$
times faster than our comparison SOP method.

The SOP methods are superior to $\alg$ and DP methods in cases where
a continuum of eigenstates is required to describe the dynamics, as
in bond-breaking, isomerization, or chemical reactions. In such cases,
$\alg$ easily becomes intractable and SOP is superior. In the large
class of energy-transfer problems, where a discrete set of eigenstates
captures the dynamics of the system, we believe $\alg$ will generally
outperform SOP methods. \diffend

\section{Example\label{sec:Example}}
\diffbegin
In this section we study a system that has the smallest Hilbert space
considered in Fig.~\ref{fig:Cost-Ratio}, in which wavefunctions
may be represented using $N_{c}=28$ terms. \diffend This small size allows
us to generate TA spectra in a matter of seconds, and therefore it
is easy to run fast parameter sweeps and map out how spectroscopic
observables change.

We explore nonlinear optical spectra for a model system presented
by Tiwari and Jonas \citep{Tiwari2018}. 
\begin{figure}
\includegraphics[width=0.5\textwidth]{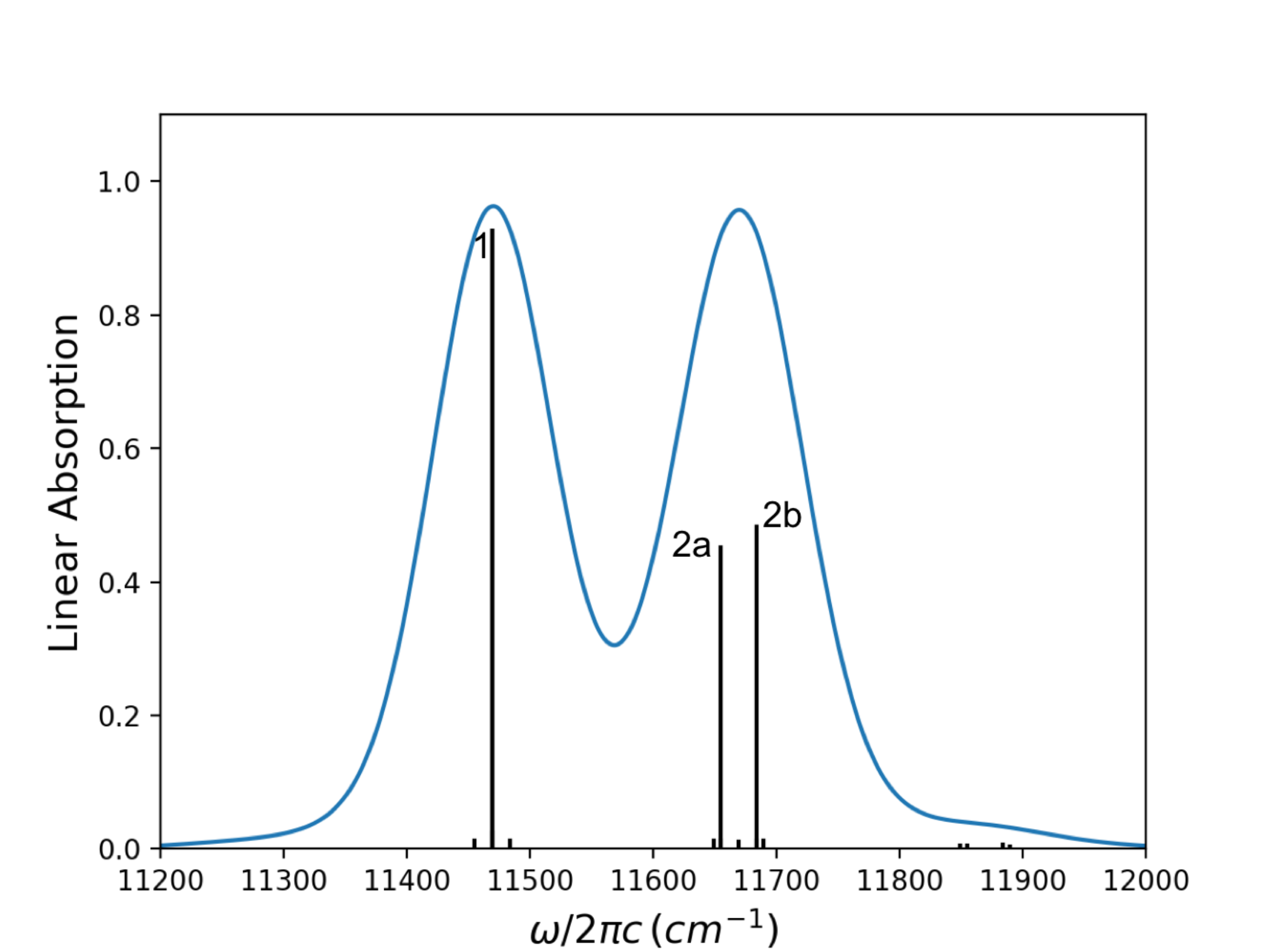}\caption{\label{fig:TiwariFig2b}Linear absorption stick spectrum calculated
using the same eigenstates as in Ref.~\onlinecite{Tiwari2018}. The
frequencies of the stick spectra are the eigenenergies of the Hamiltonian
Eq.~\ref{eq:nmer} with $k=s=2$, $\omega_{i}=200\,\text{cm}^{-1}$
and $S_{i}=0.025$ for each monomer subunit. The two subunits have
dipoles $\protect\abs{\mu_{ag}}=\protect\abs{\mu_{bg}}$ and $\protect\cvec{\mu}_{ag}\cdot\protect\cvec{\mu}_{bg}=0$.
The heights of the sticks are $\protect\abs{\hat{\protect\cvec{\mu}}_{\phi'\phi}}^{2}$.
The spectrum is thermally averaged at a temperature of 80 K, using
the 3 lowest energy ground vibrational states, $\protect\ket{00},\protect\ket{01},\protect\ket{10}$.
The blue curve represents the absorption cross-section, which is obtained
by broadening the stick spectrum with homogeneous and inhomogenous
linewidths of $10\,\text{cm}^{-1}$ and $45\,\text{cm}^{-1}$, respectively.}
\end{figure}
Their model system is inspired by the Fenna-Matthews-Olson complex
\citep{Fenna1975} and consists of a molecular dimer formed from two
electronic \diffbegin TLS \diffend, each locally coupled to a single harmonic vibrational
mode, \diffbegin which is the case of $k=s=2$ from Eq.~\ref{eq:nmer}. \diffend

The system couples to optical pulses in the electric-dipole approximation,
with transition dipole matrix elements $\cvec{\mu}_{ga}$ and $\cvec{\mu}_{gb}$.
Following Ref.~\citenum{Tiwari2018}, we consider the homodimer to
have nearly identical subunits, with identical vibrational frequencies
$\omega_{a}=\omega_{b}=200\,\text{cm}^{-1}$, \diffbegin Huang-Rhys factors $S=0.025$, \diffend
and with $E_{b}-E_{a}=150$~cm$^{-1}$. We consider that $\cvec{\mu}_{ga}$,
$\cvec{\mu}_{gb}$ have the same magnitude but extend the model by
varying the angle $\theta$ between these transition dipoles, corresponding
to varying the angle between the two subunits. Due to the small Huang-Rhys
factors considered in Ref.~\onlinecite{Tiwari2018}, only a small
number of vibrational states contribute to third-order spectra, allowing
truncation of the vibrational Hilbert space, as described in Sec.~\ref{subsec:Efficiency-Improvements}.
Reference~\onlinecite{Tiwari2018} did not calculate TA spectra;
first we compare our results for linear absorption and then present
our TA predictions.

Reference~\onlinecite{Tiwari2018} demonstrates how the Huang-Rhys
factors split one exciton peak into two vibronic peaks. We plot the
linear absorption spectrum for this system in Fig.~\ref{fig:TiwariFig2b},
choosing lineshape parameters to visually reproduce results from Ref.~\onlinecite{Tiwari2018}.
We label the peaks in Fig.~\ref{fig:TiwariFig2b} following Tiwari
and Jonas. There are two broad peaks centered around $f_{1}=11469.5\,\text{cm}^{-1}$
and $f_{2}=11669.6\,\text{cm}^{-1}$, with the latter composed of
two peaks, $2a$ and $2b$, separated by $\Omega_{1}=29\,\text{cm}{}^{-1}$.
However, this splitting is invisible to linear absorption because
peaks $2a$ and $2b$ smear together.

\begin{figure}
\includegraphics[width=1\columnwidth]{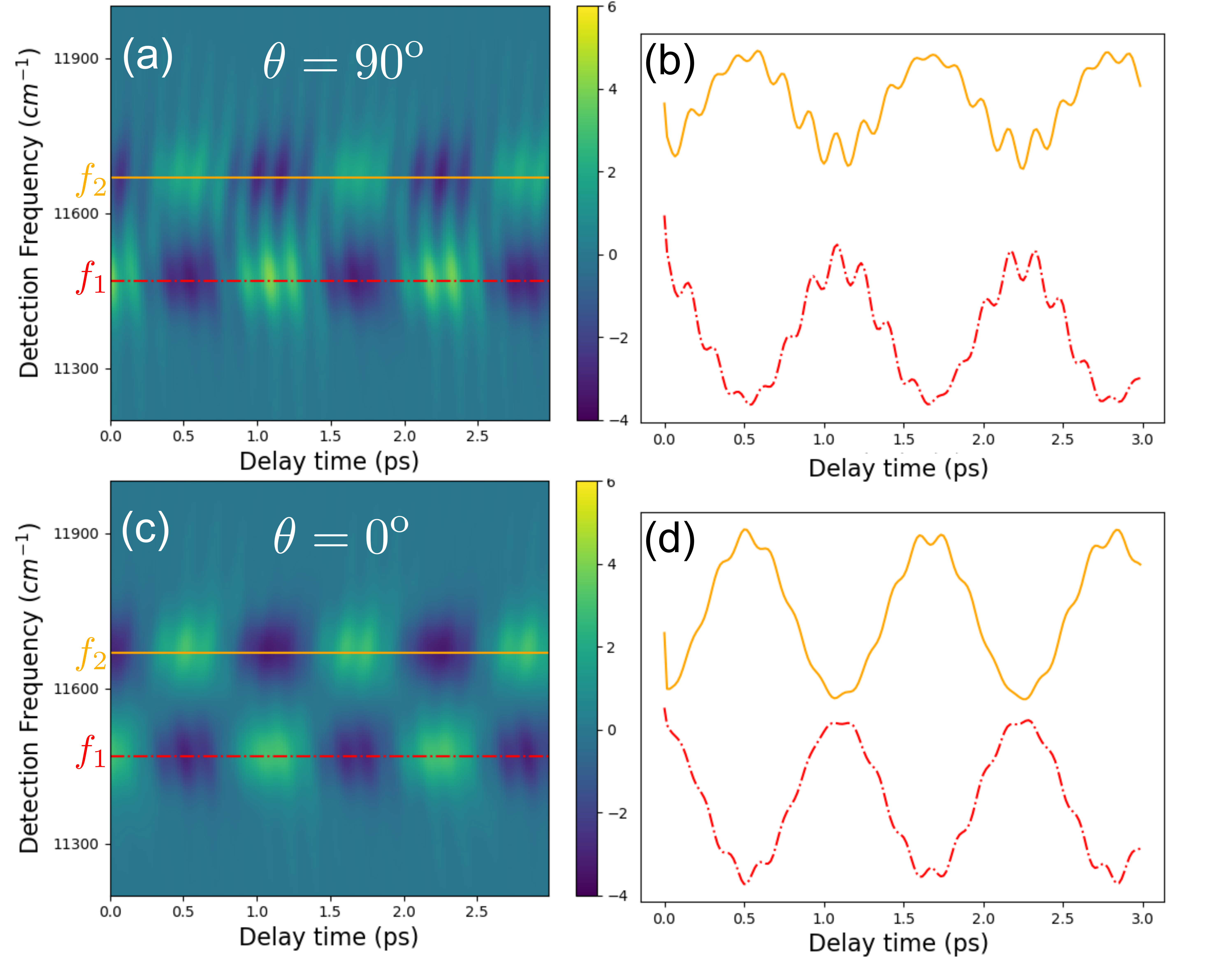}\caption{\label{fig:PumpProbeExample}Transient absorption with Gaussian pump
and probe pulses centered on $f_{2}$ and with FWHM of 12~fs. (a)
Dipole angle $\theta=90^{\text{o}}$, (b) linecuts of (a) at $f_{2}$
(orange) and $f_{1}$ (red, dash-dotted), (c) Dipole angle of $\theta=0^{\text{o}}$,
(d) linecuts of (c) at $f_{2}$ (orange) and $f_{1}$ (red, dash-dotted).
In both cases the low-frequency beats are out of phase between $f_{1}$
and $f_{2}$. The high-frequency beats are in phase between $f_{1}$
and $f_{2}$ for $\theta=90^{\text{o}}$ and are reduced in amplitude
and out of phase for $\theta=0^{\text{o}}$.}
\end{figure}

$\Omega_{1}$ is, however, the dominant beat frequency in a TA experiment,
shown in Fig.~\ref{fig:PumpProbeExample}, corresponding to oscillations
repeating about every 1.15~ps. The TA signal is detected as two broad
features centered at $f_{1}$ and $f_{2}$. Notably, the oscillations
centered around $f_{1}$ are roughly $90^{\text{o}}$ out of phase
with those at $f_{2}$, and there is a node halfway between them.
This type of nodal feature is widely observed and has been studied
in many vibrational systems \citep{Liebel2015,McClure2014,Cina2016}.

\diffbegin As mentioned in Section II.E, to achieve convergence of these TA spectra
to within 1\%, it is sufficient to have $N_{f,DEM}=N_{f,GSM}=15$
and $N_{f,SEM}=30$, while $N_{c,DEM}=N_{c,GSM}=10$ and $N_{c,SEM}=8$. \diffend
Each TA spectrum is isotropically averaged, with the same homogeneous
and inhomogeneous linewidths in the linear absorption spectrum. Each
TA spectrum, which is of the form $S_{TA}(T,\omega)$, consists of \diffbegin
$m_{T}=180$ delay times, and $M_{t}=286$, 562, 1391 detection frequency
points for FWHM pulse durations of 62, 31 and 12~fs, respectively,
and takes about 3 seconds to generate on a 2.3GHz Macbook Pro. The
equivalent calculation using the SOP code (see Section~\ref{subsec:Split-operator})
would take almost 6 days to converge to the same accuracy if run on
the same machine. This system is the leftmost point of Fig.~\ref{fig:Cost-Ratio},
and the RKE method would take over 7 minutes to produce each spectrum. \diffend
\begin{figure}
\includegraphics[width=1\columnwidth]{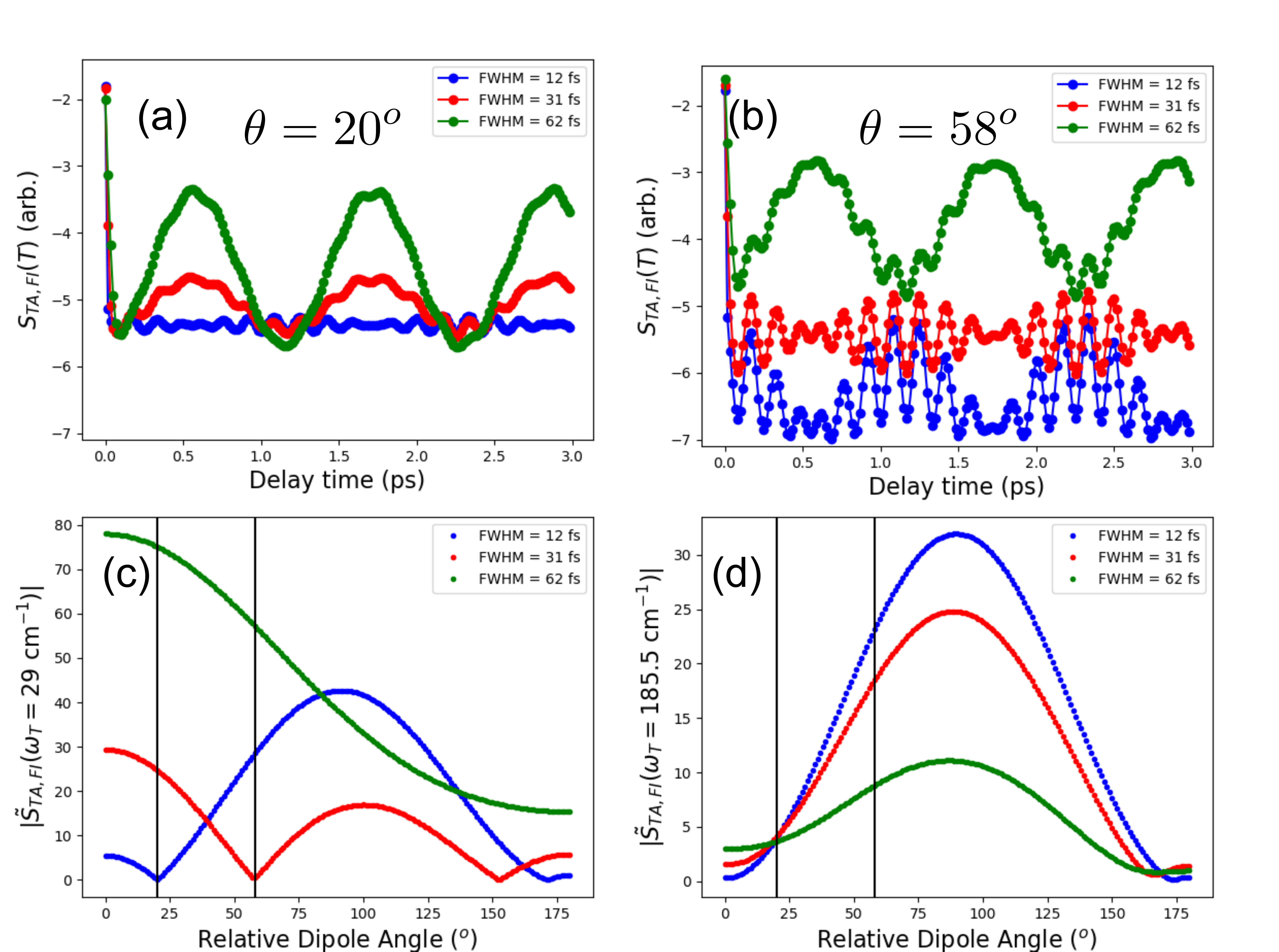}\caption{\label{fig:TAFI}Demonstration of parameter variations for the same
system as in Fig.~\ref{fig:PumpProbeExample}. Panels (a) and (b)
show frequency-integrated transient absorption (FITA) spectra for
(a) $\theta=20^{\text{o}}$ and (b) $\theta=58^{\text{o}}$ with three
different Gaussian pulse durations. There are two prominent time scales
present: the $\Omega_{1}$ beat (corresponding to the 1.15~ps oscillation)
and the $\Omega_{2}$ and $\Omega_{3}$ beats (corresponding to the
180~fs and 156~fs oscillations). Panels (c) and (d) show the magnitude
of these beats of the FITA spectra at (c) $\omega_{T}=\Omega_{1}$
and (d) $\omega_{T}=\Omega_{2}$ as a function of $\theta$. Black
vertical lines indicate $\theta$ from panels (a) and (b). Note that
at $\omega_{T}=\Omega_{1}$, the FITA magnitude shows a dramatic dependence
on both dipole angle and electric field shape, demonstrating the importance
of taking into account electric field shapes. The signals at $\omega_{T}=\Omega_{2}$
are less sensitive. }
\end{figure}

The efficiency of $\alg$ enables rapid study of a wide range of parameters.
The frequency-integrated TA signal, \diffbegin
\begin{equation}
S_{FI,TA}(T)=\int\d\omega S_{TA}(T,\omega),\label{eq:FITA}
\end{equation} \diffend
is particularly sensitive to electric field shape \citep{Yuen-Zhou2012,Johnson2014}.
Figure~\ref{fig:TAFI}(a,b) shows frequency-integrated TA spectra
with dipole angles $\theta=20^{\circ}$, $58^{\circ}$ and three optical
pulse durations. The responses with different pulse shapes are quite
different for both values of $\theta$. In order to study the differences
over a wide range of $\theta$ we study the Fourier transform of $S_{FI,TA}(T)$
with respect to the delay time:
\[
\tilde{S}_{FI,TA}(\omega_{T})=\frac{1}{\sqrt{2\pi}}\int_{-\infty}^{\infty}\d TS_{FI,TA}(T)e^{i\omega_{T}T},
\]
and track the response at $\omega_{T}=\Omega_{1}$ and $\omega_{T}=\Omega_{2}=f_{2a}-f_{1}=185.5\,\text{cm}^{-1}$
in Fig.~\ref{fig:TAFI}(c,d). A nearly impulsive pulse (12~fs FWHM
- blue curves) gives a dramatically different response as compared
to somewhat longer pulses (35~fs - red curves, 70~fs - green curves).
Notice that the longest pulse considered here is still short compared
to the fastest time scale of the system, $\Omega_{3}$, corresponding
to \diffbegin a period of \diffend 156~fs, so the large changes of the signals with pulse
duration are a warning sign that accurate study of finite-pulse effects
is important to correctly model nonlinear spectra. This study is easily
tractable using $\alg$, \diffbegin taking only 30 minutes to run, but would
have taken 3 days to run using RKE and more than a year using our
SOP implementation on the same computer. \diffend

\section{Conclusion}

$\alg$ is a computationally efficient method for calculating perturbative
nonlinear spectroscopies of systems with finite dimensional relevant
Hilbert spaces. It includes all finite-duration and pulse-overlap
effects, enabling accurate modeling of and fitting to nonlinear spectroscopic
data. The current introduction has focused on closed systems, but
$\alg$ can be readily extended to open systems with time-independent
Liouvillians, in which case its cost scales as $N_{c}^{4}$ rather
than $N_{c}^{2}$, where $N_{c}$ is the relevant Hilbert space dimension. \diffbegin
We have shown the methods that can be used to aggressively reduce
the required dimension $N_{c}$, which enables $\alg$ to be highly
efficient for small systems and competitive for surprisingly large
problems. For vibronic systems with thousands of relevant states,
$\alg$ generally outperforms our comparison DP method and is competitive
for many systems with tens of thousands of relevant states; for larger
systems, DP methods are generally superior. \diffend

The intent of the publicly available $\alg$ code is to present a
package that makes it easy to translate Feynman diagrams into code.
Since $\alg$ readily enables consideration of arbitrary pulse shapes,
the effects of pulse durations, chirps, etc.~on nonlinear spectroscopies
can now be included as a matter of course in analyses of experiments.
\diffbegin Given the large number of problems under active study with tens to
hundreds of relevant states, we hope that $\alg$ will enable easy
and rapid spectroscopic prediction and analysis including pulse shape
effects that are often neglected. \diffend

\section*{Acknowledgments}

We acknowledge helpful conversations with Joel Yuen-Zhou, Ivan Kassal,
Mark Embree,
Luc Robichaud, and Eduard Dumitrescu. We also acknowledge funding
from the Natural Sciences and Engineering Research Council of Canada
and the Ontario Trillium Scholarship.

%

\appendix

\section{Computational cost of $\protect\alg$\label{sec:Computational-Cost-Analysis}}

We derive the computational cost of $\alg$, highlighting which parameters
are required for convergence and which are at the user's discretion,
focusing on the case of a 2DPE rephasing spectrum, as in Eq.~\ref{eq:3variableSignal}.

We derive the computational cost of calculating the 2DPE rephasing
signal. We briefly outline what is required for this calculation by
working backwards from Eq.~\ref{eq:3variableSignal}. \diffbegin Symbols used
in this section are summarized in Table~\ref{tab:Summary-of-symbols-1}. \diffend
Calculation of the signal requires that we determine $\cvec P_{\text{2D}}^{(3)}(\tau,T;\omega)$
at the desired values of $\tau$ and $T$. $\alg$ directly calculates
$\cvec P_{\text{2D}}^{(3)}(t)$ for one pair of $(\tau,T)$ at a time.
We calculate $\cvec P_{\text{2D}}^{(3)}(t)$ at sufficient time points,
$M_{t}$, in order to obtain the desired frequency resolution. The
cost of the FFT to obtain $\cvec P_{\text{2D}}^{(3)}(\omega)$ from
$\cvec P_{\text{2D}}^{(3)}(t)$ is negligible compared with the other
costs of $\alg$. Assuming that we require the polarization field
at $m_{\tau}$ values of $\tau$ and $m_{T}$ values of $T$, the
cost of the full spectrum is $m_{\tau}m_{T}\text{Cost}(\cvec P_{\text{2D}}^{(3)}(t))$.
$\cvec P_{\text{2D}}^{(3)}(t)$ is a sum of three Feynman diagrams
(see Fig.~\ref{fig:Three-Rephasing-Diagrams}), each of which is
calculated separately. We focus on \diffbegin $\cvec P_{ESA}^{(3)}(t)$, which
is often the dominant cost. The cost of the other diagrams follows
directly from this derivation. \diffend

\begin{table}
\caption{\label{tab:Summary-of-symbols-1} \diffbegin Summary of symbols used to describe
computational cost of a 2DPE rephasing signal. $M$, $M_{E}$, $\epsilon$,
and $dt$ are convergence parameters. \diffend}

\begin{tabular}{cc}
 & \tabularnewline
\hline 
\hline 
Symbol & Definition\tabularnewline
\hline 
$M$ & Number of time points to resolve pulses $\alg$\tabularnewline
$M_{E}$ & Number of time points to resolve pulse in RKE\tabularnewline
$\epsilon$ & Local tolerance of RK45 algorithm\tabularnewline
$dt$ & Fixed step size for RKE and $\alg$ during pulses\tabularnewline
$dt_{RK}$ & Adaptive step size of RK45 algorithm\tabularnewline
\hline 
 & \tabularnewline
\end{tabular}
\end{table}

Using Eq.~\ref{eq:P_of_t_ESA}, the cost of $\cvec P_{ESA}^{(3)}(t)$
can be broken into three parts: calculating the two necessary perturbed
wavepackets and the cost of the dipole matrix element of those wavepackets.
This latter cost turns out to be the dominant cost asymptotically.
Since \diffbegin $\ket{\psi_{a}(t)}=K_{a}\ket{\psi^{(0)}(t)}$ and $\ket{\psi_{bc}(t)}=K_{c}K_{b}\ket{\psi^{(0)}(t)}$,
the cost of calculating these wavepackets is, at first glance, the
cost of three calls to the $K_{j^{(*)}}$ operator, which is the heart
of $\alg$. However, $\ket{\psi_{a}(t)}$ corresponds to the interaction
with the first pulse, which arrives at a fixed time, and therefore
need only be calculated once. Further, $\ket{\psi_{b}(t)}$ corresponds
to the second pulse, which arrives at $m_{\tau}$ different coherence
times $\tau$. Therefore $\ket{\psi_{b}(t)}$ only needs to be calculated
$m_{\tau}$ different times. The only wavefunction that must be recalculated
for every pair $\tau,T$ is the one caused by the third interaction,
$\ket{\psi_{bc}(t)}$. Therefore the dominant cost of calculating
the perturbative wavefunctions for many different values of $T$ is
simply one call to $K_{c}$. 

To determine the cost of $K_{j^{(*)}}$, we first explain how wavefunctions
are stored in $\alg$. \diffend Each wavefunction is represented in the same
way. For example, the second-order wavefunction is \diffbegin
\begin{equation}
\ket{\psi_{bc}(t)}=\sum_{\phi=1}^{N_{c,DEM}}e^{-i\omega_{\phi}t}c_{\phi,ac^{*}}(t)\ket{\phi},\label{eq:psi^n}
\end{equation} \diffend
where each of the $c_{\phi,ac^{*}}(t)$ are calculated at $M$ evenly
spaced time points . The spacing $dt$ is determined by the shape
of the pulse amplitude $A_{i}(t)$ (see Eq.~\ref{eq:pulse_shape}).
Recall from Eq.~\ref{eq:c_phi} that $c_{\phi,pj^{(*)}}(t)$ only
varies in the interval $(t_{j,\min},t_{j,\max})$. Outside of this
interval, $c_{\phi,pj^{(*)}}$ is constant and can thus be trivially
extended to any time points needed. This property allows wavefunctions
to be stored for re-use without a significant memory cost. For simplicity
in this discussion we assume that $t_{j,\max}-t_{j,\min}$ is the
same for each pulse. Since we are discretizing a continuous convolution
integral, $M=\frac{t_{j,\max}-t_{j,\min}}{dt}$ is a convergence parameter.
For any electric field that does not go strictly to zero, the choice
of $t_{j,\min}$ and $t_{j,\max}$ must also be checked for convergence.

Inspecting Eq.~\ref{eq:Kj}, the cost of $\hat{K}_{j}$ is \diffbegin $N_{c,X}$ \diffend
times the cost of evaluating $\theta*y_{\phi}$ to obtain $c_{\phi,pj}$.
Calculating $y_{\phi}$ is dominated by the cost of taking the dipole-weighted
sum over $\phi'$ (we are neglecting the small cost of multiplying
by the pulse amplitude $\varepsilon_{j}^{(*)}$ and by the time evolution
factors $e^{i\omega_{\phi}(t-t')}$ and $e^{-i\omega_{\phi'}(t-t')}$).
The sum has a cost of $\alpha N_{c,X}M$, where $\alpha$ is the cost
of \diffbegin multiplying \diffend two complex numbers. We calculate $\theta*y_{\phi}$
using the convolution theorem and the FFT. Since we are interested
in the linear convolution, we evaluate $\theta$ at $2M-1$ points
and zero-pad $y_{\phi}$ to be size $2M-1$. We only calculate the
FFT of $\theta$ once, making that cost negligible. We calculate the
FFT of $y_{\phi}$, which has a cost of $\beta'(2M-1)\log_{2}\left(2M-1\right)\approx2\beta'M\log_{2}M+O(M)$,
where $\beta'$ depends upon the implementation of the FFT.\footnote{We find that when calculating the convolution between a step-function
$\theta(t)$ and another function $y(t)$, we achieve convergence
much more quickly when we use an odd number of time points, with half
positive, half negative, and the point $t=0$ with associated value
$\theta(0)=0.5$. The FFT is fastest for arrays of length $2^{k}$,
which is impossible in this case. To keep $\beta'$ small, we use
the FFTW library and pick a size $2M-1$ that has only small prime-factors.} We multiply the two ($\tilde{y}_{\phi}$ and $\tilde{\theta}$),
which has cost $O(M)$, and then take the inverse FFT of the product
(cost $2\beta'M\log_{2}M+O(M)$). Thus the total cost of the convolution
is $4\beta'M\log_{2}M+O(M)$. The cost to obtain one coefficient $c_{\phi,pj^{(*)}}(t)$
to highest order in $M$ and $N$ is then\textbf{ }
\begin{equation}
\text{Cost}(c_{\phi,pj^{(*)}}(t))=\alpha N_{c,X}M+\beta M\log_{2}M+O(M),\label{eq:Costcnphiphip}
\end{equation}
where $\beta=4\beta'$. The cost of calling $K_{j^{(*)}}$ is then \diffbegin
$N_{c,Y}\text{Cost}(c_{\phi,pj^{(*)}})$, where $N_{c,X}$ is the
dimension of the old manifold and $N_{c,Y}$ is the dimension of the
new manifold, giving 
\begin{equation}
\text{Cost}(K_{j^{(*)}})=N_{c,Y}M(\alpha N_{c,X}+\beta\log_{2}M)+O(MN_{c,Y}),\label{eq:CostKB}
\end{equation}
and so $\text{Cost}(K_{j^{(*)}})\sim MN_{c,X}N_{c,Y}$ for large $N_c$.
In the case of $\ket{\psi_{bc}(t)}$, then, we have a cost $\sim MN_{c,SEM}N_{c,DEM}$.
Given that $N_{c,DEM}$ is often the largest value of $N_{c,X},X\in\{GSM,SEM,DEM\}$,
the ESA diagram often dominates the cost of $\alg$. The other diagrams
involve wavefunctions that move between the GSM and the SEM, and therefore
the cost of each $K_{j}$ scales as $N_{c,GSM}N_{c,SEM}$. \diffend

Once we know each of the necessary wavefunctions at its $M$ time
points, we evolve the wavefunctions using $\hat{U}_{0}$ to include
$M_{t}$ times points, in order to obtain the desired frequency resolution
of the final spectrum. The $M_{t}$ points are spaced by the same
$dt$, and span from just before the last pulse arrives, $t_{c,\min}$,
until the signal has decayed to an appropriate cut-off due to the
optical dephasing described in Section~\ref{subsec:Including-Optical-Dephasing}.
This evolution is of negligible cost since we know the exact form
of $\hat{U}_{0}(t)$.

We then calculate the expectation value $\boldsymbol{P}_{ESA}^{(3)}(t)=\bra{\psi_{a}(t)}\hat{\boldsymbol{\mu}}\ket{\psi_{bc}(t)}$,
which has a cost of \diffbegin $\alpha N_{c,SEM}N_{c,DEM}(M+M_{t})$, since the
polarization field must capture both the pulse turn-on described by
$M$ points, and the decay of the field described by the additional
$M_{t}$ points, so in total $\text{Cost}(\cvec P_{ESA}^{(3)}(t))=\alpha N_{c,SEM}N_{c,DEM}(M+M_{t})+\text{Cost}(K_{j})$.
\diffend The signal must be calculated at $m_{\tau}$ $\tau$ points and $m_{T}$
$T$ points, so we arrive at the full cost of the \diffbegin ESA signal
\begin{align}
\text{Cost}(S_{ESA}(\tau,T,\omega)) & =m_{\tau}m_{T}\bigg(\alpha N_{c,SEM}N_{c,DEM}(M+M_{t})\nonumber \\
 & +N_{c,DEM}M\big(\alpha N_{c,SEM}+\beta\log_{2}M\big)\bigg)\label{eq:Total_SE_Cost}
\end{align}
to highest order in $N_{c,X}$, $M_{t}$ and $M$. In general we find that
$M<M_{t}$, and the cost of evaluating the necessary wavefunctions
is sometimes less than half of the cost of obtaining the total signal. \diffend
In Section~\ref{sec:Example}, we calculate the TA signal, which
is composed of four Feyman diagrams with costs less than or equal
to that in Eq.~\ref{eq:Total_SE_Cost}, with $m_{\tau}=1$, since
the TA signal has $\tau=0$. In Section~\ref{sec:Example} we use \diffbegin
$M=21$ and $M_{t}\approx1000$, in order to achieve convergence of
better than $1\%$.\footnote{The value of $M_{t}$ depends upon the pulse shape and the optical
dephasing $\gamma$. We resolve the polarization field from $t_{c,\min}$
until $t_{c}+6.91/\gamma$.}

Assuming that $M<M_{t}$, and that $\alpha N_{c}\gg\beta\log_{2}M$,\footnote{In our implementation of this algorithm, $\beta\log_{2}M/\alpha\approx100$.}
the total cost scales as 
\begin{equation}
C_{\alg}(S_{ESA})\sim m_{\tau}m_{T}N_{c,SEM}N_{c,DEM}M_{t}.\label{eq:UF2-ESA-cost}
\end{equation}
\diffend

Note that $\alg$ assumes that the eigenvalues and eigenstates of
$\hat{H}_{0}$ have already been attained. If the dimension of $\hat{H}_{0}$
is $N_{f}$, then the computational cost of that diagonalization can
scale as $N_{f}^{3}$, which can exceed the cost of $\alg$ itself,
especially since $\alg$ reduces the relevant Hilbert space dimension
$N_{c}$ so aggressively. For many systems, however, $\hat{H_{0}}$
is sparse, allowing efficient iterative methods to be used to find
its eigenvalues and eigenstates. The scaling of those algorithms is
beyond the scope of this manuscript, but we simply note that there
are many important cases where solving $\hat{H}_{0}$ is not computationally
limiting. For example, vibronic systems of coupled chromophores, each
with local harmonic oscillators, are extremely sparse, and an efficient
algorithm to determine their eigenstates will be detailed separately.

\diffbegin

\section{RKE implementation and scaling}

\subsection{Implementation\label{sec:RK45Implementation}}

Direct propagation (DP) methods solve the Schrodinger equation 
\[
\frac{d\ket{\psi^{(n)}(t)}}{dt}=-iH_{0}\ket{\psi^{(n)}(t)}-iH'_{j}(t)\ket{\psi^{(n-1)}(t)},
\]
where we have set $\hbar=1$, by propagating the ODE forward in time
domain.

In order to compare the cost of $\alg$ to the cost of DP methods,
we have implemented a hybrid DP method that uses the adaptive-step
size 4-5$^{\text{th}}$ order Runge-Kutta ODE solver from scipy (called
RK45) to propagate the system dynamics
\[
\frac{d\ket{\psi^{(n)}}}{dt}=-iH_{0}\ket{\psi^{(n)}},
\]
and uses the first-order Euler method to include the perturbation
\[
-iH'_{j}(t)\ket{\psi^{(n-1)}(t)},
\]
with a previously calculated lower-order wavepacket $\ket{\psi^{(n-1)}(t)}$,
as described in Refs.~\onlinecite{Engel1991,Yuen-Zhou2014}. As with
$\alg$, we assume that each pulse $\cvec E_{j}$ is non-zero only
during times $t_{j,\text{min}}<t<t_{j,\text{max}}$. When all of the
pulses are zero, propagation is simply handled by RK45. Given a local
absolute and relative tolerance $\epsilon$, RK45 can integrate from
$t_{i}$ to $t_{f}$ using an adaptively determined step size $dt_{RK}$.
The wavefunctions are stored at equally spaced time points with time
step $dt$, which must be short compared to the optical pulses and
in our implementation we set to the same $dt$ used in $\alg$. This
fixed time step is used for the Euler method and allows the reuse
of lower-order wavefunctions for many values of pulse delay times.
Note that as mentioned at the end of Section \ref{subsec:UF2-vs-RKE},
this choice of $dt$ does not converge the RKE spectra to the same
accuracy as $\alg$.\footnote{For problems with a small number of delay times, it may be more efficient
to use the RK45 solver for both system and perturbation propagation,
but this choice would make wavefunction reuse more complicated. As
this manuscript is interested in the case where many different experimental
configurations are considered, we have chosen an implementation that
we believe gives the best performance, though we do not claim to have
optimized all aspects.} Given a state vector $\ket{\psi(t_{i})}$, we use the notation $\ket{\psi(t_{i}+dt)}=U_{0}(dt)\ket{\psi(t_{i})}$
to represent the RK45 propagation.

To include the perturbation $H'_{j}(t)$, we begin by defining the
first-order wavefunction $\ket{\psi_{a}(t<t_{a,\text{min}})}=0$.
We define $\ket{\psi_{a}(t_{a,\text{min}})}=-iH_{a}'(t_{a,\text{min}})dt\ket{\psi^{(0)}(t_{a,\text{min}})}$,
where $\ket{\psi^{(0)}(t_{a,\text{min}})}$ is a stationary state
(possibly with an evolving phase factor). We propagate until $t_{a,\text{max}}$
in steps of size $dt=(t_{a,\text{max}}-t_{a,\text{min}})/M_{E}$.
Let $m$ run from 0 to $M_{E}$. Then 
\begin{multline*}
\ket{\psi_{a}(t_{a,\text{min}}+mdt)}=U_{0}(dt)\ket{\psi_{a}(t_{a,\text{min}}+(m-1)dt)}\\
+i\cvec{\mu}\cdot\cvec e_{a}\varepsilon_{a}(t_{a,\text{min}}+mdt)\ket{\psi^{(0)}(t_{a,\text{min}}+mdt)}.
\end{multline*}
Once we have obtained $\ket{\psi_{a}(t_{a,\text{max}})}$, we obtain
$\ket{\psi_{a}(t>t_{a,\text{max}})},$ using RK45 alone. This method
works in general for the $n^{th}$ order wavefunction generated by
an interaction with the $j^{th}$ pulse. Given the $\left(n-1\right)^{\text{st}}$
wavefunction $\ket{\psi_{p}(t)}$, we define $\ket{\psi_{pj^{(*)}}(t_{j,\text{min}})}=i\cvec{\mu}\cdot\cvec e_{j}^{(*)}\varepsilon_{j}^{(*)}(t_{j,\text{min}})\ket{\psi_{p}(t_{j,\text{min}})}$,
and then at all times until $t_{j,\text{max}}$ using
\begin{multline*}
\ket{\psi_{pj^{(*)}}(t_{a,\text{min}}+mdt)}=U_{0}(dt)\ket{\psi_{pj^{(*)}}(t_{a,\text{min}}+(m-1)dt)}\\
+i\cvec{\mu}\cdot\cvec e_{j}^{(*)}\varepsilon_{j}^{(*)}(t_{a,\text{min}}+mdt)\ket{\psi_{p}(t_{a,\text{min}}+mdt)}.
\end{multline*}
Once again, for all times after $t_{j,\text{max}}$, we only use RK45
to propagate the wavefunction in its manifold.

\subsection{Computational cost analysis\label{subsec:RK45Cost}}

We derive the computational cost of calculating the 2DPE rephasing
signal, now using the RK45-Euler hybrid (RKE) method outlined in Section~\ref{sec:RK45Implementation},
following a similar path as our derivation of the cost of $\alg$,
working backwards from Eq.~\ref{eq:3variableSignal}. Symbols used
in this section are summarized in Tables~\ref{tab:Summary-of-symbols}-\ref{tab:Summary-of-symbols-1}.
As for $\alg$, we begin by considering the ESA diagram, and derive
the cost of $\cvec P_{ESA}^{(3)}(t)$.

The cost of RKE is the cost of obtaining the necessary wavefunctions,
$\ket{\psi_{a}(t)}$ and $\ket{\psi_{bc}(t)}$, and computing $\cvec P_{ESA}(t)=\bra{\psi_{a}(t)}\cvec{\mu}\ket{\psi_{bc}(t)}$.
In the Condon approximation, $\cvec{\mu}$ is a sparse rectangular
matrix consisting of diagonal blocks, so this expectation value has
a cost of $\alpha qr_{\mu}N_{f,DEM}$. Even outside of the Condon
approximation, $\cvec{\mu}$ is still sparse, and therefore the cost
is still $O(N_{f})$; the number of entries per row, $r_{\mu}$, is
simply a little larger.

The remaining cost is the cost of obtaining the necessary wavefunctions.
The cost of obtaining a perturbative wavefunction breaks down into
two parts: the cost of the RK45 algorithm to propagate $H_{0}$ and
the cost of the Euler method to include the perturbation $H_{j}'$.
We first derive the cost of RK45 for times after $t>t_{j,\text{max}}$,
where RK45 can take full advantage of its adaptive step size. In this
regime, the user sets a local tolerance $\epsilon$, and RK45 determines
what step size $dt_{RK}$ satisfies that tolerance, and we approximate
it as a constant. To take each step, RK45 must make 6 function evaluations
of the form $H_{0}\ket{\psi}$ and sum the results with the required
weights. Each evaluation has the cost of a sparse matrix-vector multiplication,
$\alpha qrN_{n,X}$, where $r$ is the average number of non-zero
entries of $H_{0}$ per row and $q$ is an additional overhead factor
for sparse matrix operations. Ignoring the extra costs when steps
are rejected, we thus find that the cost to take a step $dt_{RK}$
is $6\alpha(qr+1)N_{f}\approx6\alpha qrN_{f,X}$. Assuming we need
to know the wavefunction at some final time $t_{f}>t_{j,\text{max}}$,
the cost of propagating the wavefunction when the pulse is off,
\begin{align*}
C_{RK} & =6\alpha qrN_{f,X}M_{RK},
\end{align*}
where $M_{RK}=(t_{f}-t_{j,\text{max}})/dt_{RK}$ is the required number
of steps after the pulse ends. This wavefunction can be relatively
inexpensively evaluated on a mesh with spacing $dt$ using the RK45
interpolator, but the computational cost is set by the adaptively
optimized $dt_{RK}$.

During the time when the pulse is non-zero, the cost of the RK45 portion
of the evolution is
\[
6\alpha qrN_{n,X}\max\left(M_{E},(t_{j,\text{max}}-t_{j,\text{min}})/dt_{RK}\right),
\]
where $M_{E}=(t_{j,max}-t_{j,min})/dt$. If $dt$ is smaller than
$dt_{RK}$, we must pay an additional cost of taking smaller steps
than field-free evolution using RK45 would require. We must also pay
the cost of adding in the perturbation. Again, since $\cvec{\mu}$
is a sparse matrix with $r_{\mu}$ entries per row, the cost of $\cvec{\mu}\ket{\psi}$
is $\alpha qr_{\mu}N_{f}$. The cost of evaluating $\varepsilon_{j}$
at a single time point is negligible, and so the cost of adding in
the perturbation is $\alpha qr_{\mu}N_{n,X}M_{E}.$ Assuming that
$dt<dt_{RK}$, we can easily add these costs to obtain the cost of
propagating when the pulse is on,
\begin{align*}
C_{E} & =\alpha q(6r+r_{\mu})N_{n,X}M_{E}.
\end{align*}
 Thus the cost of calculating a single wavefunction is 
\begin{align*}
C_{RKE}\left(\ket{\psi_{pj^{(*)}}}\right) & =C_{E}+C_{RK}
\end{align*}
Despite the fact that $dt$ is generally smaller than $dt_{RK}$,
it is still often the case that $M_{RK}>M_{E}$. In general we find
that $M_{E}=100$ is usually sufficient to resolve the pulse interaction
and converge the spectroscopic signals to within 1\%. In contrast,
we often find that $M_{RK}>100$, and therefore the cost of obtaining
a perturbative wavefunction is usually dominated by the propagation
cost in the absence of the electric field. This observation depends
upon the imposed decay of the polarization field, $\gamma$, as described
in Section \ref{subsec:Including-Optical-Dephasing}. As for $\alg$,
once the polarization field is obtained, we multiply by $e^{-\gamma|t-t_{probe}|}$,
and therefore we generally need to resolve the wavefunction to $t_{f}=t_{\text{probe}}+\frac{5}{\gamma}$.

We developed this hybrid method for the purposes of making a fair
comparison with $\alg$. We assert that if $C_{E}>C_{RK}$, there
is likely a more efficient algorithm, for example using RK45 both
while the pulse is on and when the pulse is off. For the purposes
of comparison, we make sure to operate in a regime where $C_{E}<C_{RK}$,
and therefore we approximate $C_{RKE}\approx C_{RK}$.

This process must be repeated 3 times in order to calculate the ESA
at a single pair of pulse delays $\tau,T$. However, $\ket{\psi_{a}}$
only needs to be calculated once, and can be stored and re-used because
we are working with a regularly spaced time grid. Similarly, $\ket{\psi_{b}}$
only needs to be calculated once for each value of $\tau$ but can
be reused for all values of $T$. The only wavefunction that must
be recalculated each time is $\ket{\psi_{bc}}$, and therefore this
cost dominates the cost of obtaining $\cvec P_{ESA}$. Therefore 
\begin{align}
C_{RKE}(S_{ESA}) & \approx m_{\tau}m_{T}C_{RK}\label{eq:RKE-ESA-cost}
\end{align}

Note that the costs of the other diagrams contributing to $\cvec P^{(3)}(t)$
have a similar form, but scale as $N_{f,SEM}$ or $N_{f,GSM}$. Since
$N_{f,SEM},N_{f,GSM}<N_{f,DEM}$ in general, the cost of obtaining
the rephasing signal is dominated by the cost of $\cvec P_{ESA}$.
\diffend
\end{document}